\documentclass[preprint]{elsarticle}

\usepackage{lineno}
\modulolinenumbers[1]

\usepackage{lipsum}
\usepackage{rotating}
\graphicspath{{figures/}}
\usepackage{pstricks, pst-node, psfrag}
\usepackage{amssymb,amsmath,amsbsy}
\usepackage{mathtools}
\usepackage{verbatim,enumerate}
\usepackage{rotating, lscape}
\usepackage{setspace}
\usepackage[hang, flushmargin]{footmisc}
\usepackage{subfig}
\usepackage{caption}
\usepackage{cancel}
\usepackage{float}
\usepackage{ragged2e}
\usepackage{centernot}
\usepackage{tikz}
\usepackage{tikzsymbols}
\usetikzlibrary{shapes,arrows,positioning,plotmarks}
\usepackage{textcomp} 
\usepackage{gensymb} 
\usepackage{upgreek}
\usepackage{verbatim}

\tikzstyle{endpt} = [rectangle, draw, fill=red!20,
text width=9.8em, text centered, rounded corners, minimum height=4em]
\tikzstyle{block} = [rectangle, draw, top color=white, bottom color=blue!20,
text width=9.8em, text centered, rounded corners, minimum height=4em]
\tikzstyle{line} = [draw, -latex', very thick]

\usepackage{listings}
\usepackage{color} 
\definecolor{mygreen}{RGB}{28,172,0} 
\definecolor{mylilas}{RGB}{170,55,241}
\definecolor{mygray}{rgb}{0.5,0.5,0.5}
\definecolor{mycyan}{RGB}{0,255,255}
\definecolor{magenta}{rgb}{1,0,1}

\definecolor{backgreen}{rgb}{0.00, 0.169, 0.212}
\definecolor{textgray}{rgb}{0.514, 0.580, 0.589}

\usepackage[T1]{fontenc}
\newcommand*\patchAmsMathEnvironmentForLineno[1]{%
\expandafter\let\csname old#1\expandafter\endcsname\csname #1\endcsname
\expandafter\let\csname oldend#1\expandafter\endcsname\csname end#1\endcsname
\renewenvironment{#1}%
{\linenomath\csname old#1\endcsname}%
{\csname oldend#1\endcsname\endlinenomath}}%
\newcommand*\patchBothAmsMathEnvironmentsForLineno[1]{%
\patchAmsMathEnvironmentForLineno{#1}%
\patchAmsMathEnvironmentForLineno{#1*}}%
\AtBeginDocument{%
\patchBothAmsMathEnvironmentsForLineno{equation}%
\patchBothAmsMathEnvironmentsForLineno{align}%
\patchBothAmsMathEnvironmentsForLineno{flalign}%
\patchBothAmsMathEnvironmentsForLineno{alignat}%
\patchBothAmsMathEnvironmentsForLineno{gather}%
\patchBothAmsMathEnvironmentsForLineno{multline}%
}

\newlength{\dhatheight}

\renewcommand{\tilde}{\widetilde}
\renewcommand{\hat}{\widehat}

\newcommand{\norm}[1]{\left\lVert#1\right\rVert}

\newcommand{\defeq}{\coloneqq}
\DeclareMathOperator\erf{erf}

\DeclareFontFamily{U}{mathx}{\hyphenchar\font45}
\DeclareFontShape{U}{mathx}{m}{n}{
	<5> <6> <7> <8> <9> <10>
	<10.95> <12> <14.4> <17.28> <20.74> <24.88>
	mathx10
}{}
\DeclareSymbolFont{mathx}{U}{mathx}{m}{n}
\DeclareFontSubstitution{U}{mathx}{m}{n}
\DeclareMathAccent{\widecheck}{0}{mathx}{"71}
\DeclareMathAccent{\wideparen}{0}{mathx}{"75}

\journal{Advances in Water Resources}

\begin{document}

\begin{frontmatter}

\title{Particle Density Estimation with Grid-Projected Adaptive Kernels\tnoteref{mytitlenote}}
\tnotetext[mytitlenote]{This work was partially supported by the Spanish Ministry of Economy and Competitiveness through project WE-NEED, PCIN-2015-248.}

\author{Guillem Sole-Mari\fnref{barca,h2ogeo}}
\ead{guillem.sole.mari@upc.edu}
\author{Diogo Bolster\fnref{nd}}
\ead{dbolster@nd.edu}
\author{Daniel Fern\`andez-Garcia\fnref{barca,h2ogeo}}
\ead{daniel.fernandez.g@upc.edu}
\author{Xavier Sanchez-Vila\fnref{barca,h2ogeo}}
\ead{xavier.sanchez-vila@upc.edu}
\fntext[barca]{Department of Civil and Environmental Engineering (DECA), Universitat Polit\`ecnica de Catalunya, Barcelona, Spain}
\fntext[h2ogeo]{Hydrogeology Group (GHS), UPC-CSIC, Barcelona, Spain}
\fntext[nd]{Department of Civil and Environmental Engineering and Earth Sciences, University of Notre Dame, South Bend, IN, USA}

\begin{abstract}

The reconstruction of smooth density fields from scattered data points is a procedure that has multiple applications in a variety of disciplines, including Lagrangian (particle-based) models of solute transport in fluids. In random walk particle tracking (RWPT) simulations, particle density is directly linked to solute concentrations, which is normally the main variable of interest, not just for visualization and post-processing of the results, but also for the computation of non-linear processes, such as chemical reactions. Previous works have shown the superiority of kernel density estimation (KDE) over other methods such as binning, in terms of its ability to accurately estimate the ``true'' particle density relying on a limited amount of information. Here, we develop a grid-projected KDE methodology to determine particle densities by applying kernel smoothing on a pilot binning; this may be seen as a ``hybrid'' approach between binning and KDE. The kernel bandwidth is optimized locally. Through simple implementation examples, we elucidate several appealing aspects of the proposed approach, including its computational efficiency and the possibility to account for typical boundary conditions, which would otherwise be cumbersome in conventional KDE.

\end{abstract}

\begin{keyword}
Particle Density
\sep
Adaptive Kernels
\sep
Random Walk Particle Tracking
\sep
Concentration Estimation
\sep
Reactive Transport
\end{keyword}

\end{frontmatter}


\section{Introduction} 
\label{sec:introduction}

Random Walk Particle Tracking (RWPT) methods are a family of methods commonly used in the hydrologic sciences to simulate transport. They are appealing as they can accurately emulate many different physical processes that occur in porous media such as diffusion, hydrodynamic dispersion, mass transfer across multiple porosity systems and linear sorption \cite{Salamon2006a,Salamon2006}. They are also conducive to simulating anomalous non-Fickian transport that arises due to medium heterogeneities below the scale of resolution \cite[e.g.][]{Berkowitz2006}. With RWPTs, the solute mass is discretized into a large number of discrete particles that move across the porous medium following deterministic and probabilistic rules, which account for the processes of advection, dispersion, matrix diffusion, etc. Lagrangian methods for simulating scalar transport, among which RWPTs are some of the most common, have been shown to be particularly useful when modeling transport in advection-dominated systems, where Eulerian methods can suffer from numerical dispersion and instabilities \cite{Salamon2006a,Benson2017}.
 
However, the main shortcoming of RWPT methods is that, without modification, they can result in very noisy concentration fields due to subsampling effects associated with the finite number of particles in the system. This can be particularly troublesome when simulating solute transport in systems where processes are governed by nonlinearities or tight coupling and interactions between different solute concentrations, of which nonlinear chemical reactions are a prime example. Linear processes such as simple degradation, slow sorption or chain reactions can efficiently be incorporated to the RWPT algorithm by means of additional probabilistic rules with little additional computational cost. On the other hand, nonlinear reactions involve interactions between neighboring particles, which adds complexity to the problem as particles do not just need to know where they are, but where all others are also, which in naive implementations would mean an $\mathcal{O}(N^2)$ numerical cost. Even with more optimized approaches that use better search algorithms \cite[e.g.][]{engdahl_ddc}, the additional numerical cost can become significant for high particle numbers.

A problem that clearly highlights these issues  and has received considerable recent attention in the literature is the simulation of bimolecular reactions of the type $\mathrm{A}+\mathrm{B}\rightarrow\mathrm{C}$ via RWPT \cite{Benson2008,Paster2013,Bolster2016,Lazaro2019}. In many such cases it has been shown that the noise associated with the particles can fundamentally change the large scale behavior of the system; in some instances, it may reflect a true noise in the system \cite{Paster2014,Ding2017}, but in others it may be a numerical artifact that leads to incorrect predictions \cite{Rahbaralam2015}. Most attempts to simulate other, more complex reactions from a Lagrangian perspective have chosen to attribute concentrations to particles instead of fixed masses \cite{Benson2016,Engdahl2017}, and to represent non-advective processes by means of mass transfer, thus allowing arbitrarily complex reactive processes to be simulated on-particle. This approach, however, hinders some of the intrinsic advantages of RWPT. For instance, species-dependent transport properties cannot be readily implemented, at least not in a quick and straightforward manner. Moreover, since solute mass tends to occupy new fluid as the simulation advances, empty particles need to be included in all those areas where it is anticipated that the solute may reach by dispersion \cite{Herrera2009}.

A first attempt at simulating arbitrarily complex kinetic chemical reactions with traditional RWPT methods was recently presented by \cite{Sole-Mari2017}. The method smooths concentration fields using an optimal kernel density estimator. In the paper, the authors derive an expression for the probability of reaction of a particle for any kinetic rate expression, using the optimal kernel density estimator as the particle support volume. This was extended in \cite{Sole2018}, where using a locally adaptive optimal kernel to determine the solute concentrations and ultimately the particle reaction probabilities were demonstrated to be superior over methods based on simple binning, where concentrations are computed by defining fixed representative volumes or bins over which the contained particle mass is assumed to be uniformly distributed. Despite potential shortcomings, the definition of a spatial discretization as a set of fixed bins is very convenient; it allows establishing links between the Lagrangian representation of the solute mass, and other properties that may be defined in space, thus readily enabling nonlinear and coupled interactions. Moreover, particle counting in a regular mesh can be very efficient from a computational viewpoint, typically much more so than a kernel density approach. Thus, one has to weigh disadvantages and benefits when deciding which approach to take. Ideally, one would have a method designed to obtain the best of all worlds. 

Another challenge that often arises in the use of RWPT is the application of nontrivial boundary conditions, best defined in an Eulerian framework. Several methods have been proposed in the recent literature to incorporate different kinds of boundary conditions to RWPT, allowing us to simulate impermeable, Dirichlet \cite{Szymczak2003}, or Robin \cite{Szymczak2004,Koch2014,Boccardo2018} boundary conditions. However, current kernel methods for the reconstruction of the concentrations \cite[e.g.,][]{Fernandez-Garcia2011} do not address the subject of boundary conditions. 

In this paper we propose a robust methodology to obtain solute concentrations from particle positions in RWPT simulations. These concentration fields can then be used to visualize the results, and perhaps more importantly, to incorporate nonlinear fate and transport processes. Although we focus on advection-dispersion in relation to a porous medium, this methodology is broadly applicable to other similar transport processes and systems. In fact, any particle-based approach that relies on some form of particle density estimation could benefit from it.
Our proposed approach can be seen as a hybrid technique, where the nonreactive part of the transport (advection-dispersion) is simulated following classical principles of RWPT, and the reactive part is decoupled from the former and assisted by a grid on which the concentrations are estimated following an improved form of the local optimal kernel density estimation technique introduced in \cite{Sole2018}. This new on-grid kernel-based method combines the practical efficiency of binning techniques with the accuracy gains of kernel methods, while also allowing us to impose Neumann, Dirichlet and Robin boundary conditions to the density estimation.

The paper is structured as follows. First, in $\S$\ref{sec:ongrid}, we describe the methodology by which we can adapt the kernel density estimation procedure described in \cite{Sole2018} to the case of a gridded domain. Then, in $\S$\ref{sec:bounds}, we explain how to correct the resulting concentration estimations to account for a variety of boundary conditions typically considered in porous media. Then, in $\S$\ref{sec:comput} we perform several computational experiments to test and also illustrate the proposed methodology, and in $\S$\ref{sec:conc} we finish with a summary and conclusions.

\section{On-Grid Concentration Estimation From Particle Positions With Local Optimal Kernels} 
\label{sec:ongrid}
Let us consider a numerical particle cloud, made up of $N$ particles of identical mass $m$, that represent the spatial distribution of the total mass of a given chemical compound (solute), over a $d$-dimensional continuum. Particle positions are given by $\left\{\mathbf{X}_1,\dots,\mathbf{X}_N\right\}$. The concentration of the compound at position $\mathbf{x}\equiv\left[x_1,\dots,x_d\right]^\mathrm{T}$ would then be 
\begin{equation}\label{c}
c\left(\mathbf x\right)=\frac{m\rho\left(\mathbf x\right)}{\phi\left(\mathbf x\right)},
\end{equation}
where $\rho\left(\mathbf{x}\right)$ is the particle density (particles per unit volume of medium), and $\phi\left(\mathbf x\right)$ is the volumetric content of the fluid per unit volume of medium. The particle density $\rho\left(\mathbf{x}\right)$ needs to be estimated from particle positions, and one method to do so is via kernel density estimation (KDE) such that
\begin{equation}\label{density}
\rho\left(\mathbf x\right)\defeq\sum_{p=1}^NW\left(\mathbf x-\mathbf X_p;\mathbf h_p\right),
\end{equation}
where $W$ is the kernel, chosen here as a ``product'' multi-Gaussian:
\begin{equation}\label{W}
W\left(\mathbf{r};\mathbf{h}\right)=\prod_{i=1}^{d}\frac{1}{\sqrt{2\pi}h_i}\exp{\left(-\frac{r_i^2}{2h_i^2}\right)},
\end{equation}
with  $\mathbf{h}\equiv\left[h_1,\dots,h_d\right]^\mathrm{T}$ being the vector of directional kernel bandwidths. Note that in \eqref{density}, every particle $p$ can have a different bandwidth $\mathbf h_p$. Expression \eqref{density} has been used in previous works \cite{Fernandez-Garcia2011,Pedretti2013,Rahbaralam2015,Sole-Mari2017,Sole2018} to link particle positions to solute concentrations. Recently, we \cite{Sole2018} proposed a technique to determine the optimal bandwidth $\mathbf h_p$ based on the minimization of the root mean squared error (RMSE) on a local environment. Here, we adapt the density estimation and the bandwidth optimization methods to apply them within a regular grid.

\subsection{Concentration in a Bin}
\label{subsec:estimation}
\begin{figure}[t]%
	\centering
	\includegraphics[width=1\textwidth]{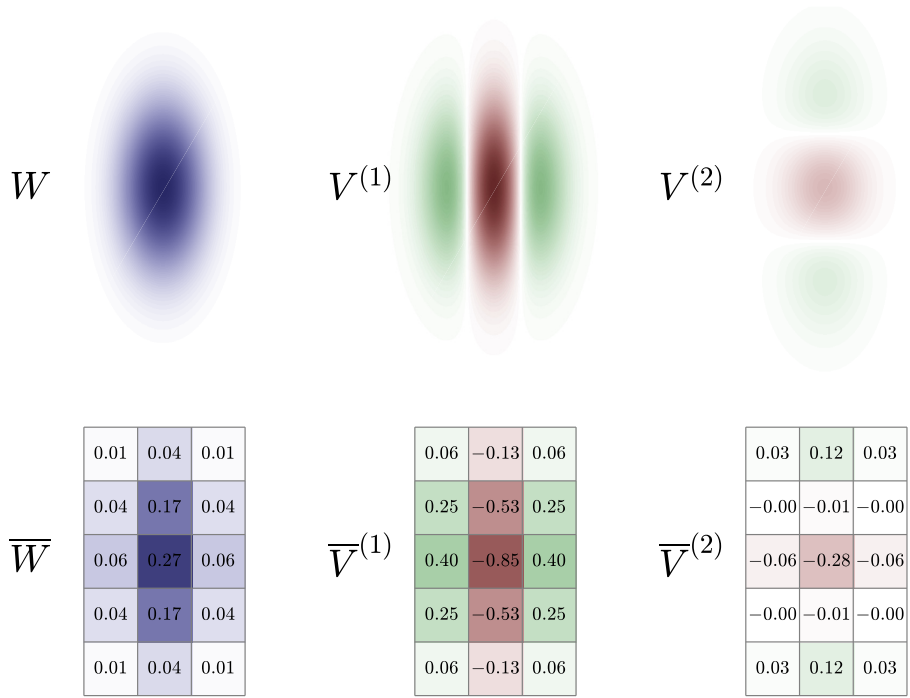}
	\caption{Graphical example of 2D kernel functions (top) and their discrete equivalents obtained by projection on a Eulerian grid (bottom). Color intensity is proportional to the kernel density, and numbers correspond to the integral over a cell ($\Lambda\rho_u$).}
	\label{fig:kernelmatrices}
\end{figure}

Let us discretize our entire domain of interest into $\nu$ regular bins of size $\boldsymbol{\uplambda}\equiv\left[\lambda_1,\dots,\lambda_d\right]^\mathrm{T}$, labeled as $u=1,\dots,\nu$. Let us also group particle positions into the centers of the containing bins; i.e., if particle $p$ falls into bin $u$, then $\mathbf X_p\approx \mathbf x_u$, where $\mathbf x_u$ is the position of the center of bin $u$. Then we can define a discrete (mean) value of the particle density $\rho$ in the $u$th bin as
\begin{equation}\label{f_disc}
\rho_u\defeq\frac1{\Lambda}\int_{\mathbf x_u-\boldsymbol{\uplambda}/2}^{\mathbf x_u+\boldsymbol{\uplambda}/2}\rho\left(\mathbf x\right)\mathrm d\mathbf x\approx\frac1{\Lambda}\sum_{\omega=1}^\nu \mu_\omega\overline W\left(\mathbf x_\omega-\mathbf x_u;\mathbf h_\omega,\boldsymbol{\uplambda}\right),
\end{equation}
where $\Lambda= \prod_{i=1}^{d}\lambda_i$ is the size of the bin, $\mu_\omega$ is the particle count (number of particles) in bin $\omega$, $\mathbf h_\omega$ is its associated kernel bandwidth, and
\begin{equation}\label{disc_int}
\overline W\left(\mathbf r;\mathbf h,\boldsymbol{\uplambda}\right)\defeq\int_{\mathbf r-\boldsymbol{\uplambda}/2}^{\mathbf r+\boldsymbol{\uplambda}/2}W\left(\mathbf r^\prime;\mathbf h\right)\mathrm d\mathbf r^\prime.
\end{equation}
The kernel $\overline W$ is a projected form of $W$. By combining \eqref{W} and \eqref{disc_int} we obtain the closed form:
\begin{equation}\label{Wprime}
\overline W\left(\mathbf r;\mathbf h,\boldsymbol{\uplambda}\right)=\frac1{2^d}\prod_{i=1}^d\left[\erf{\left(\frac{r_i+\lambda_i/2}{\sqrt 2h_i}\right)}-\erf{\left(\frac{r_i-\lambda_i/2}{\sqrt 2h_i}\right)}\right].
\end{equation}
Note from \eqref{f_disc} that, since all $\mathbf{x}_u$, $\mathbf{x}_\omega$ belong to a regular grid of size $\boldsymbol{\uplambda}$, then $\mathbf{r}$ in \eqref{Wprime} can be written as
\begin{equation}\label{disc}
\mathbf{r}=\boldsymbol{\uplambda}\odot\mathbf{z},
\end{equation}
where $\mathbf{z}=\left[z_1,\dots,z_d\right]^\mathrm{T}$ is a vector of integers ($z_i\in\mathbb Z$, $\forall i=1,\dots,d$), and $\odot$ is the Hadamard (element-wise) product. That is, $\mathbf{z}$ is a cell index, with $\mathbf{z}=\mathbf{0}$ corresponding to the cell where the kernel is centered. Then expression \eqref{Wprime} can be rewritten as
\begin{equation}\label{Wprime_lam}
\overline W\left(\boldsymbol{\uplambda}\odot\mathbf{z};\mathbf h,\boldsymbol{\uplambda}\right)=\frac1{2^d}\prod_{i=1}^d\left[\erf{\left(\lambda_i\frac{z_i+1/2}{\sqrt 2h_i}\right)}-\erf{\left(\lambda_i\frac{z_i-1/2}{\sqrt 2h_i}\right)}\right].
\end{equation}
By examining equation \eqref{Wprime_lam}, we see that for any given ratio $\mathbf h\oslash\boldsymbol{\uplambda}$ (where $\oslash$ is the Hadamard division) the set of required evaluations of $\overline W$ in \eqref{f_disc} (setting a cutoff distance) can be fully defined as a $d$-dimensional matrix (see leftmost illustrations in Figure \ref{fig:kernelmatrices}). Then, by limiting the possible values that $\mathbf{h}$ can adopt to a discrete set, it is possible to greatly speed up evaluation of \eqref{f_disc} by storing these matrices after their first computation, hence never having to re-evaluate \eqref{Wprime} for similar values of $\mathbf{h}_\omega$. More details on the properties, generation, cut-off correction, storage and use of matrix kernels can be found in \ref{app:matrices}.

Now, the kernel-based evaluation of particle densities on bins consists of two steps. First, a simple binning is performed to obtain the particle count $\mu_\omega$ in every bin, which can be seen as an initial, perhaps noisy, ``pilot'' density estimation. Then, smoothing \eqref{f_disc} is performed using the matrix kernels corresponding to bandwidths $\mathbf h_\omega$ to obtain densities $\rho_u$.

Finally, the concentration in bin $u$ can be calculated as
\begin{equation}\label{c_disc}
c_u\defeq\frac1{\Lambda}\int_{\mathbf x_u-\boldsymbol{\uplambda}/2}^{\mathbf x_u+\boldsymbol{\uplambda}/2}c\left(\mathbf x\right)\mathrm d\mathbf x\approx\frac{m\rho_u}{\phi_u},
\end{equation}
where $\phi_u\defeq\phi\left(\mathbf{x}_u\right)$.
Details on the local selection of parameter $\mathbf h_\omega$ to use in \eqref{f_disc} are given in the following section.

\subsection{The Optimal Kernel Bandwidth $\mathbf{h}$}\label{sub:optimal_h}
Let us decompose the bandwidth $\mathbf{h}$ by defining the bandwidth scale $\hat h$ and the vector of ``shape'' parameters $\mathbf{s}\equiv\left[s_1,\dots,s_d\right]^\mathrm{T}$:
\begin{equation}\label{h}
	\hat h\defeq\left(\prod_{i=1}^dh_i\right)^{\frac1d},\qquad s_i\defeq h_i/{\hat h},
\end{equation}
i.e., $\mathbf{h}\equiv\hat h\mathbf{s}$, with $\prod_{i=1}^ds_i=1$.
Then, following \cite{Sole2018}, the local optimal bandwidth scale is
\begin{equation}\label{h_opt_loc}
	\hat h_u=\left[\frac{d \ n_u}{\left(4\pi\right)^{\frac{d}{2}}T_u}\right]^{\frac{1}{d+4}},
\end{equation}
where
\begin{equation}\label{n_disc}
n_u\defeq\int \rho\left(\mathbf x\right)W\left(\mathbf{x}-\mathbf{x}_u;\boldsymbol{\upsigma}_u\right)\mathrm d\mathbf x\approx \sum_{\omega=1}^\nu\rho_\omega\overline W\left(\mathbf x_\omega-\mathbf x_u;\boldsymbol{\upsigma}_u,\boldsymbol{\uplambda}\right),
\end{equation}
and
\begin{equation}\label{T}
T_u=
\begin{cases}
\varPsi_{11,u}, &\mathrm{for}\ d=1, \\[10 pt]
2\left(\varPsi_{11,u}\varPsi_{22,u}\right)^\frac12+2\varPsi_{12,u}, &\mathrm{for}\ d=2,\\
3\left(\varPsi_{11,u}\varPsi_{22,u}\varPsi_{33,u}\right)^\frac13+\sum_{i\neq j\neq k}\varPsi_{ij,u}\left(\dfrac{\varPsi_{kk,u}^2}{\varPsi_{ii,u}\varPsi_{jj,u}}\right)^\frac16, &\mathrm{for}\ d=3.
\end{cases}
\end{equation}
In \eqref{T}, the functionals $\varPsi_{ij,u}$ are defined as
\begin{equation}\label{psi_disc}
\begin{split}
\varPsi_{ij,u}\defeq\int \kappa^{(i)}\left(\mathbf{x}\right)\kappa^{(j)}\left(\mathbf{x}\right)W&\left(\mathbf{x}-\mathbf{x}_u;\boldsymbol{\upsigma}_u\right)\mathrm d\mathbf x \\
&\approx\sum_{\omega=1}^\nu\kappa^{(i)}_\omega\kappa^{(j)}_\omega\overline W\left(\mathbf x_\omega-\mathbf x_u;\boldsymbol{\upsigma}_u,\boldsymbol{\uplambda}\right),
\end{split}
\end{equation}
where
\begin{equation}\label{kappa}
\kappa^{(i)}\left(\mathbf{x}\right)\defeq\frac{\partial^2\rho}{\partial x_i^2}\left(\mathbf{x}\right)=\sum_{p=1}^NV^{(i)}\left(\mathbf x-\mathbf X_p;\mathbf g_p^{\left(i\right)}\right),
\end{equation}
with the kernel function $V^{(i)}$ being defined as
\begin{equation}\label{gaussian_deriv}
V^{(i)}\left(\mathbf r;\mathbf g\right)\defeq\frac{\partial^2 W}{\partial r_i^2}=\left(\frac{r_i^2}{g_i^4}-\frac1{g_i^2}\right)W\left(\mathbf r;\mathbf g\right).
\end{equation}
In \eqref{psi_disc}, $\kappa^{(i)}_u$ is defined as the mean value of $\kappa^{(i)}$ in bin $u$:
\begin{equation}\label{kappa_disc}
\kappa^{(i)}_u\defeq
\frac1{\Lambda}\int_{\mathbf x_u-\boldsymbol{\uplambda}/2}^{\mathbf x_u+\boldsymbol{\uplambda}/2}\kappa^{(i)}\left(\mathbf x\right)\mathrm d\mathbf x
\approx\frac{1}{\Lambda}\sum_{\omega=1}^\nu \mu_\omega\overline V^{(i)}\left(\mathbf x_\omega-\mathbf x_u;\mathbf g^{(i)}_\omega,\boldsymbol{\uplambda}\right),
\end{equation}
with
\begin{equation}\label{disc_int_V}
\overline V^{(i)}\left(\mathbf r;\mathbf g,\boldsymbol{\uplambda}\right)\defeq\int_{\mathbf r-\boldsymbol{\uplambda}/2}^{\mathbf r+\boldsymbol{\uplambda}/2}V^{(i)}\left(\mathbf r^\prime;\mathbf g\right)\mathrm d\mathbf r^\prime.
\end{equation}
The notation $\mathbf{g}_\omega^{\left(i\right)}$ \emph{in lieu} of $\mathbf{h}_\omega$ indicates that this bandwidth can adopt a different value than that of kernel $W$, and also different values for different directions of derivative. By combining \eqref{gaussian_deriv} and \eqref{disc_int_V} we obtain the following closed form:
\begin{align}\label{Vhat}
&\overline V^{(i)}\left(\mathbf{r};\mathbf{g},\boldsymbol{\uplambda}\right)=-\frac{1}{2^{(d-\frac12)}\sqrt{\pi}g_i^3}\times \nonumber \\
&\qquad\left[\left(r_i+\frac{\lambda_i}2\right)\exp{\left(-\frac{\left(r_i+\lambda_i/2\right)^2}{2g_i^2}\right)}-\left(r_i-\frac{\lambda_i}2\right)\exp{\left(-\frac{\left(r_i-\lambda_i/2\right)^2}{2g_i^2}\right)}\right]\times \nonumber \\
&\qquad\qquad\prod_{j\neq i}\left[\erf{\left(\frac{r_j+\lambda_j/2}{\sqrt 2g_j}\right)}-\erf{\left(\frac{r_j-\lambda_j/2}{\sqrt 2g_j}\right)}\right].
\end{align} 
Similar to $\overline W$, the quantity $\lambda_i^2\overline V^{(i)}$ can be stored as a matrix (Figure \ref{fig:kernelmatrices}) with values that only depend on the ratio $\mathbf{g}\oslash\boldsymbol{\uplambda}$.  Additional corrections are performed on kernel $\overline V^{(i)}$ to ensure that it keeps the main properties of the original kernel $V^{(i)}$ despite the projection. Details on these corrections, as well as the generation, storage and use of matrix kernels can be found in \ref{app:matrices}.

The shape vector is also determined by the ``roughness'' functionals $\varPsi_{ij,u}$:
\begin{equation}\label{s}
s_{i,u}=\left(\dfrac{\hat\varPsi_u}{\varPsi_{ii,u}}\right)^{\frac14},\qquad \hat\varPsi_u\defeq\left(\prod_{j=1}^{d}\varPsi_{jj,u}\right)^{\frac1d}.
\end{equation}
Detailed information on the theory behind these expressions can be found in \cite{Sole2018}. Nevertheless, here we give an intuitive explanation: In \eqref{h_opt_loc}, $n_u$ is a smooth average of the particle density over a local environment, whereas $T_u$, termed as ``effective roughness'', is a measure of the square of the second spatial derivatives of the densities, also averaged over a local environment. Thus, typically, those areas where particle densities are higher, or where they form ``peaks'' and ``valleys'' that are more pronounced, will yield smaller kernel bandwidths (lower values of $\hat h_u$). In \eqref{s}, we see that the elongation of the kernel bandwidth in a direction $i$ will be inversely correlated to the (region-averaged) squared derivatives of the density in that direction, normalized by its geometric average over all dimensions; that is, the kernel will tend to stretch along the direction of minimum curvature. In sum, the local kernel size and shape adapts to ``mimic'' the features of the surrounding particle cloud. 

At this juncture, the computation of the optimal bandwidth $\mathbf{h}_u$ to use in \eqref{f_disc} requires the input of two additional kernel bandwidths: $\boldsymbol{\upsigma}_u$ in \eqref{n_disc} and \eqref{psi_disc} ; and $\mathbf{g}_{\omega}^{\left(i\right)}$ in \eqref{kappa_disc}. These are addressed in the following two sections, respectively.

\subsection{The Integration Support $\boldsymbol{\upsigma}$}\label{sub:sigma}
Computation of \eqref{n_disc} and \eqref{psi_disc} requires the definition of a Gaussian integration support, represented here by its vector of directional bandwidths $\boldsymbol{\upsigma}_u$. In the original development of the local optimization methodology \cite{Sole2018}, an isotropic, spatially constant support $\sigma_i\equiv\hat\sigma$, $\forall i=1,\dots,d$ was proposed, such that $\hat\sigma=3\hat h^\mathcal{G}$, with $\hat h^\mathcal{G}$ defined as the global AMISE-optimal kernel bandwidth scale. Not only is this approach completely heuristic in nature, but it also renders the local kernel indirectly dependent on a global feature, compromising local benefits. Here, we overcome this problem by introducing the concept of an equivalent normal particle distribution. We assume that the local optimal kernel scale $\hat h_u$ is also the global optimal kernel bandwidth $\hat h_u^\sigma$ associated with a virtual Gaussian distribution of variance $\hat\sigma_u^2$, composed of a virtual number of particles $N_u^\sigma$; the classic expression for the AMISE-optimal kernel bandwidth in this case is \cite{Silverman1986}:
\begin{equation}\label{Silverman}
	\hat h_u\equiv \hat h_u^\sigma=\left[\frac{4}{\left(d+2\right)N_u^\sigma}\right]^{\frac1{d+4}}\hat\sigma_u.
\end{equation}
We then impose that this virtual distribution has locally matching values with the actual particle distribution for $\rho_u$ and $n_u$, the latter being defined as in \eqref{n_disc}, using $\boldsymbol{\upsigma}_u$ as the integration support. Then, it can be shown (see \ref{app:22}) that the following relation holds, 
\begin{equation}\label{sigma_rel_2}
N_u^\sigma=\frac{\left(\sqrt{8\pi}\hat\sigma_u\right)^d n_u^2}{\rho_u}.
\end{equation}
Combining \eqref{Silverman} with \eqref{sigma_rel_2}, and after some algebraic manipulation, we obtain:
\begin{equation}\label{sigma}
\hat\sigma_u=\left[\frac{\left(d+2\right)\left(8\pi\right)^{\frac d2}n_u^2\hat h_u^{d+4}}{4\rho_u}\right]^{\frac14}.
\end{equation}
The integration support used in \eqref{n_disc} and \eqref{psi_disc} is then $\boldsymbol{\upsigma}_u=\hat{\sigma}_u\mathbf{1}$, where $\mathbf{1}$ is a $d\times 1$ vector of ones. Note that expression \eqref{sigma} is recursive, in the sense that $h_u$, $n_u$ and $\rho_u$ are all affected by some previous choice of $\hat\sigma_u$. Nonetheless, it can be implemented iteratively. More details on the iterative implementation are given in $\S$\ref{sub:optimization}.

\subsection{The Curvature Kernel Bandwidth $\mathbf{g}^{\left(i\right)}$}\label{sub:g}
Computation of the optimal bandwidth as presented in $\S$\ref{sub:optimal_h} requires a bandwidth for the estimation of the particle density curvatures ($\mathbf g^{(i)}_\omega$ in \eqref{kappa}). Previously \cite{Sole2018}, the Improved Sheather-Jones plug-in method by \emph{Botev et al.} \cite{botev2010} was used to determine this bandwidth. Here, we rely on the equivalent Gaussian particle distribution described in $\S$\ref{sub:sigma} to determine a local value for $\mathbf g^{(i)}_\omega$ recursively based on $\mathbf{h}_\omega$. 

Above, we have defined both these bandwidths as diagonal. Here we further assume that $\mathbf g^{(i)}$ is isotropic, 
\begin{equation}\label{g}
	g^{(i)}_j\equiv\hat g^{(i)}, \ \forall j=1,\dots,d.
\end{equation}
We also assume that the anisotropy of $\boldsymbol{\upsigma}_u$ is identical to that of bandwidth $\mathbf{h}_u$ (represented by vector $\mathbf{s}_u$). Then, in a similar fashion to what was done in $\S$\ref{sub:sigma} for $\hat h_u$, the value of bandwidth scale $\hat g_u^{(i)}$ is assumed to match its $\mathrm{AMISE}$-optimal magnitude within the virtual Gaussian distribution of particles defined in $\S$\ref{sub:sigma} through $\boldsymbol{\upsigma}_u$ and $N_u^\sigma$. This magnitude (see derivation in \ref{app:gopt}) is
\begin{equation}\label{gopt}
	\hat g_u^{(i)}\equiv\hat g_u^{(i),\sigma}=\left[\frac{4+2^{\frac{d}{2}+4}}{3\left(d+4\right)N_u^\sigma}\right]^{\frac1{d+6}}\hat\sigma_u \vartheta_i\left(\mathbf{s}_u\right),
\end{equation}
where $\vartheta_i$ is a function of the anisotropy of $\mathbf{s}_u$, with a value of 1 in the isotropic case (i.e., $s_i=1, \ \forall i=1,\dots,d$): 
\begin{equation}\label{theta}
	\vartheta_i\left(\mathbf{s}\right)=\left[\frac{1}{d+4}\sum_{j=1}^{d}\frac{1+4\delta_{ij}}{s_i^4s_j^2}\right]^{-\frac{1}{d+6}},
\end{equation}
where $\delta_{ij}$ is the Kronecker delta. 
By combining \eqref{gopt} and \eqref{Silverman}, we can determine the ratios between optimal bandwidth scales:
\begin{equation}\label{gamma_i}
\gamma^{(i)}_u\defeq\frac{\hat g_u^{(i)}}{\hat h_u}=\alpha\cdot\left(N_u^\sigma\right)^\beta\cdot\vartheta_i\left(\mathbf{s}_u\right),
\end{equation}
\begin{equation}\label{a&b}
\alpha\defeq \left[\frac{1+2^{\frac{d+4}{2}}}{3\cdot2^{\frac{4}{d+4}}}\right]^{\frac{1}{d+6}}\cdot\frac{\left(d+2\right)^{\frac1{d+4}}}{\left(d+4\right)^{\frac1{d+6}}},\qquad \beta\defeq \frac{2}{\left(d+4\right)\left(d+6\right)},
\end{equation}
with $N_u^\sigma$ obtained from \eqref{sigma_rel_2}. It is worth noting that $\alpha$ in \eqref{gamma_i} is a constant value close to 1 (ranging from $1.04$ for $d=1$ to $1.12$ for $d=3$), and $\beta\ll1$ (meaning that the ratio $\gamma^{(i)}_u$ is very rigid with respect to $N_u^\sigma$). Hence $\hat g_u^{(i)}$ is typically just somewhat larger than $\hat h_u$ for a wide range of values of $N_u^\sigma$. For instance, in 2D, assuming isotropy ($\vartheta_i=1$), $\gamma^{(i)}_u\approx1.31$ for $N_u^\sigma=10^2$, and $\gamma^{(i)}_u\approx1.75$ for $N_u^\sigma=10^5$. Using definition \eqref{gamma_i}, we can recursively obtain $\hat g_u^{(i)}=\gamma^{(i)}_u \hat h_u$, and then, in equation \eqref{kappa_disc}, we use $\mathbf{g}_u=\hat g_u \mathbf{1}$. 

\subsection{Optimization Algorithm}\label{sub:optimization}

The approach that we propose to determine the locally optimal bandwidth can be seen as a fixed-point iteration method: A set of local kernel bandwidths is given as an input, and a new set is obtained as an output, until convergence. Below, for any variable ``$a$'', the notation $\{a\}$ indicates the set of all $a_u$. Before starting the iteration process, the particle counts $\{\mu\}$ are computed, and, if unavailable, an initial pilot $\{\hat \sigma\}$ is defined such that $\hat \sigma_u=3\hat h_u$. Then, the structure of one iteration can be summarized as follows:

\begin{enumerate}
	\item Compute $\{\rho\}$ by \eqref{f_disc} using the input $\{\mathbf{h}\}$.
	\item To obtain $\{\hat \sigma\}$ and $\{n\}$:
		\subitem Compute pilot $\{n\}$ by \eqref{n_disc} using $\{\rho\}$ and $\{\hat \sigma\}$.
		\subitem Compute $\{\hat \sigma\}$ by \eqref{sigma} using $\{\hat h\}$, $\{\rho\}$ and pilot $\{n\}$.
		\subitem Compute $\{n\}$ by \eqref{n_disc} using $\{\rho\}$ and $\{\hat \sigma\}$.
	\item For $i=1,\dots,d$:
		\subitem Compute $\{{\hat g}^{(i)}\}$ by \eqref{gamma_i} using $\{\rho\}$, $\{n\}$, $\{\hat \sigma\}$ and $\{\mathbf{s}\}$.
		\subitem Compute $\{\kappa^{(i)}\}$ by \eqref{kappa_disc} using $\{g^{(i)}\}$.
	\item For $i=1,\dots,d$, for $j=i,\dots,d$:
		\subitem Compute $\{\varPsi_{ij}\}$ by \eqref{psi_disc} using $\{\hat \sigma\}$, $\{\kappa^{(i)}\}$ and $\{\kappa^{(j)}\}$.
	\item Compute the new $\{\mathbf{h}\}$ by \eqref{h_opt_loc} and \eqref{s} using $\{n\}$ and all $\{\varPsi_{ij}\}$.
\end{enumerate}

The iteration process may be exited when relative changes in $\{\mathbf{h}\}$ are below some tolerance level. In the context of a full RWPT simulation, since changes in the optimal kernel bandwidth typically occur at a much slower pace than solute transport \cite{Sole2018}, the optimization does not need to be performed at every time step of the transport time discretization, and if performed often enough, a single iteration may suffice. In between optimizations, particles should ``carry'' the identifier of the bin where they were located at the time of the latest optimization; then, the bandwidth at a bin is the average of the bandwidths associated to the contained particles. 

\section{Boundary Condition Corrections}\label{sec:bounds}

Kernel methods typically suffer from boundary bias problems \cite[e.g.,][]{Marron_bound}. Near no flux boundaries, the standard KDE as written in \eqref{density} may for instance produce an underestimation bias; as its support can cross the boundary in question, it may unphysically project nonzero solute concentrations on the other side of such boundaries, where no such mass can actually exist. Similar problems would occur for constant concentration or fixed flux boundaries, where without some correction, kernels will project incorrect and unphysical concentrations at and across the boundaries. Such problems also arise with the proposed discrete estimator in \eqref{f_disc}. However, for the discrete case, the position of the bins with respect to the boundaries is known and does not change over the course of a simulation. This allows us to efficiently introduce corrections to account for the presence and influence of boundaries, that would otherwise be unfeasible (or at least extremely cumbersome) directly from \eqref{density}.

Here we propose an  approach based on the assumption that the Gaussian kernel that represents a particle's support volume behaves like a purely diffusive process, whose interactions with boundaries are well understood. In other words, we treat the kernel associated with a particle as if it were the result of a very fast diffusive process that has resulted in the spatial spreading of the particle's mass, with an initial condition of a Dirac delta located at the particle's physical position. If the particle is far from the boundary, this results in the standard Gaussian kernel \eqref{W}. Previous works \cite[e.g.,][]{botev2010} acknowledge this link between KDE and diffusion. The correction on the kernel that we propose is independent of the implementation of the boundary condition itself on the RWPT algorithm \cite[see][]{Szymczak2003}; that is, the particles undergoing the random walks in the system are unaware of the kernel and so their motion must still account for the presence of a boundary (e.g. reflection on a no flux boundary). Note also that, in principle, our proposed approach could be extended to any kernel-based Lagrangian method \cite[e.g.,][]{Sole2019SPH} in a bounded domain with physical boundary conditions. For most commonly used  boundary conditions, we derive the simple reflection principles that can be used to modify the kernel near a regular boundary that is aligned with the principal directions of the kernel. Then, we also propose an approach to extend this procedure to the case of irregularly-shaped boundaries, which present unique challenges. 

\subsection{Regular Boundaries}\label{sub:regular}
All boundary conditions most typically used in transport simulations can be written in terms of a balance of mass fluxes between the inner ($+$) and the outer ($-$) side of the boundary; i.e.:
\begin{equation}\label{Inlet_balance}
	\mathbf{n}^\mathrm{T}\left[\left(\phi\mathbf{D}\boldsymbol \nabla c\right)^+-\left(\phi\mathbf{D}\boldsymbol \nabla c\right)^-\right]=\mathbf{n}^\mathrm{T}\left[\left(\mathbf{q}c\right)^+-\left(\mathbf{q}c\right)^-\right],
\end{equation}
where $\mathbf{n}$ is the unit vector normal to the boundary. Depending on the assumptions made for the outer side of the boundary, this balance can result in a variety of common boundary conditions.

\subsubsection{Impermeable/Outlet}\label{sub:noflux}

If the boundary is assumed to be a no flux boundary (impermeable), we set $\mathbf{n}^\mathrm{T}\mathbf{q}=0$, and $\left(\phi\mathbf{D}\boldsymbol \nabla c\right)^-=0$. In this case, equation \eqref{Inlet_balance} becomes the homogeneous Neumann boundary condition, which, for boundary $\mathfrak{N}$, can be written  as
\begin{equation}\label{Neumann}
	\mathbf{n}^\mathrm{T}\left(\mathbf{x}_\mathfrak{N}\right)\boldsymbol{\nabla}c\left(\mathbf{x}_\mathfrak{N}\right)=0,\qquad \mathbf{x}_\mathfrak{N}\in \mathfrak{N}.
\end{equation}
Note that \eqref{Neumann} would also be obtained at an outlet, by assuming $\mathbf{q}c^-=\mathbf{q}c^+$, and $\left(\phi\mathbf{D}\boldsymbol \nabla c\right)^-=0$. 

Let us consider a particle $p$ located at $\mathbf{X}_p$. 
The density kernel associated with this particle is $W\left(\mathbf{x}-\mathbf{X}_p;\mathbf{h}_p\right)$, if we neglect the effect of the boundary.  As noted above, let us assume that this density distribution can be seen as the Green's function of a virtual, fast diffusive process. Then, if the boundary is regular and aligned with the principal directions of the kernel, the kernel is altered by the proximity to $\mathfrak{N}$ by a reflection principle such that
\begin{equation}\label{W_N}
	W_\mathfrak{N}\left(\mathbf{x},\mathbf{X}_p;\mathbf{h}_p\right)\defeq W\left(\mathbf{x}-\mathbf{X}_p;\mathbf{h}_p\right)+W(\mathbf{x}-\tilde{\mathbf{X}}_p;{\mathbf{h}}_p),
\end{equation}
where $\tilde{\mathbf{X}}_p$ is the mirror of $\mathbf{X}_p$ through $\mathfrak{N}$:
\begin{equation}\label{mirror_p}
\tilde{\mathbf{X}}_p\defeq\mathbf{X}_p-2\mathbf{n}\mathbf{n}^\mathrm{T}\left(\mathbf{X}_p-\mathbf{x}_\mathfrak{N}\right).
\end{equation}
Because of the regularity assumption,
\begin{equation}\label{interchangeable}
	\vert\mathbf{x}-\tilde{\mathbf{X}}_p\vert\equiv\vert\tilde{\mathbf{x}}-{\mathbf{X}}_p\vert,
\end{equation}
where
\begin{equation}\label{mirror}
\tilde{\mathbf{x}}\defeq\mathbf{x}-2\mathbf{n}\mathbf{n}^\mathrm{T}\left(\mathbf{x}-\mathbf{x}_\mathfrak{N}\right),
\end{equation}
which allows the corrected density estimation (i.e. replacing $W$ with $W_\mathfrak{N}$ in \eqref{density}) to be expressed as a function of the uncorrected density estimation:
\begin{equation}\label{Nmirror}
	\rho_\mathfrak{N}\left(\mathbf{x}\right)\defeq\sum_{p=1}^NW_\mathfrak{N}\left(\mathbf x,\mathbf X_p;\mathbf h_p\right)=\rho\left(\mathbf{x}\right)+\rho\left(\tilde{\mathbf{x}}\right).
\end{equation}
Hence, densities can be computed conventionally by \eqref{density}, then the corrected density at a point is obtained by adding the density for the mirror symmetry point. It follows directly, that for the on-grid methodology presented in this work,
\begin{equation}\label{Nmirror_disc}
	\rho_{\mathfrak{N},u}=\rho_u+\rho_{\tilde u},
\end{equation}
where $\rho_{\tilde u}$ is the uncorrected density \eqref{f_disc} computed at the mirror bin $\tilde u$, whose center is located at
\begin{equation}\label{xu_mirr}
\tilde{\mathbf{x}}_u=\mathbf{x}_u-2\mathbf{n}\mathbf{n}^\mathrm{T}\left(\mathbf{x}_u-\mathbf{x}_\mathfrak{N}\right).
\end{equation}

\subsubsection{Dirichlet}\label{subsub:Dirichlet}
Assuming in \eqref{Inlet_balance} that the dispersive flux, $\phi\mathbf{D}\boldsymbol \nabla c$, is the same on the outer and inner sides of the boundary, and that the outer concentration is prescribed, we obtain a Dirichlet boundary condition,
\begin{equation}\label{key}
	c\left(\mathbf{x}_\mathfrak{D}\right)=c_\mathrm{o},
\end{equation}
where $c_\mathrm{o}$ is the prescribed concentration. In this case, the boundary can be permeable to a diffusive process. This is true in both directions, so part of a particle's mass may fall outside the domain, and mass may enter the domain as well. Based again on the assumption that the kernel represents a fast diffusive process, we can separate this process into two parts following the superposition principle, first accounting for the initial condition with a homogeneous boundary condition (a perfectly absorbing boundary), and second accounting for the inhomogeneous part: 
\begin{equation}\label{Superposition}
	\begin{split}
		\rho_\mathfrak{D}\left(\mathbf{x}\right)&=\rho_{\mathfrak{D}\mathrm{H}}\left(\mathbf{x}\right)+\rho_{\mathfrak{D}\mathrm{I}}\left(\mathbf{x}\right) \\ &\equiv\sum_{p=1}^NW_{\mathfrak{D}\mathrm{H}}\left(\mathbf x,\mathbf X_p;\mathbf h_p\right)+\sum_{p=1}^NW_{\mathfrak{D}\mathrm{I}}\left(\mathbf x,\mathbf X_p;\mathbf h_p\right).
	\end{split}
\end{equation} 
The solution to the homogenous part, $W_{\mathfrak{D}\mathrm{H}}$, given a regular boundary aligned with the principal directions of the kernel, is \cite{stochasticproblems}:
\begin{equation}\label{absorbing}
	W_{\mathfrak{D}\mathrm{H}}\left(\mathbf x,\mathbf X_p;\mathbf h_p\right)=W\left(\mathbf{x}-\mathbf{X}_p;\mathbf{h}_p\right)-W(\mathbf{x}-\tilde{\mathbf{X}}_p;{\mathbf{h}}_p).
\end{equation}
The inhomogeneous part can also be solved under the same assumptions. Constant diffusion from a regular boundary follows the 1D analytical solution in the direction normal to the boundary \cite{vanGenuchten1982}:
\begin{equation}\label{justDbound}
\rho_{\mathfrak{D}\mathrm{I}}\left(\mathbf{x};\mathbf{h}\right)=\rho_\mathrm{o}\mathrm{Erfc}\left(\frac{1}{\sqrt{2}}\mathbf{n}\left[\left(\mathbf{x}-\mathbf{x}_\mathfrak{D}\right)\oslash\mathbf{h}\right]\right),
\end{equation}
with 
\begin{equation}\label{f_o}
\rho_\mathrm{o}\defeq\frac{\phi\left(\mathbf{x}_\mathfrak{D}\right)c_\mathrm{o}}{m},
\end{equation}
assuming a constant $\phi\approx\phi\left(\mathbf{x}_\mathfrak{D}\right)$ near the boundary. Expression \eqref{justDbound} is also the Green's function of diffusion for an initial condition of uniform density $2\rho_\mathrm{o}$ at the outer side of the boundary \cite{Szymczak2003}. Thus, we note that equation \eqref{justDbound} is approximately equivalent to a sum of mirror kernels (as in \eqref{Superposition}) provided that
\begin{equation}\label{justDbound_W}
	W_{\mathfrak{D}\mathrm{I}}\left(\mathbf x,\mathbf X_p;\mathbf h_p\right)=\frac{2\rho_\mathrm{o}}{\rho_*(\mathbf{X}_p)}W(\mathbf{x}-\tilde{\mathbf{X}}_p;{\mathbf{h}}_p),
\end{equation}
where $\rho_*$ is the unknown true density. We replace it by a pilot estimate at the containing bin $\omega$, i.e.,
\begin{equation}\label{approximation_fo}
	W_{\mathfrak{D}\mathrm{I}}\left(\mathbf x,\mathbf X_p;\mathbf h_p\right)\approx\frac{2\mu_\mathrm{o}}{\mu_\omega}W(\mathbf{x}-\tilde{\mathbf{X}}_p;{\mathbf{h}}_p),\qquad \mathbf{X}_p\approx \mathbf{x}_\omega,
\end{equation}
with $\mu_\mathrm{o}\defeq\Lambda \rho_\mathrm{o}$. Then, substituting \eqref{absorbing} and \eqref{approximation_fo} into \eqref{Superposition}, and integrating as in \eqref{f_disc}, the on-grid corrected density estimate becomes:
\begin{equation}\label{Dmirror_disc}
	\rho_{\mathfrak{D},u}=\rho_u+\rho_{\tilde u}^{\mathrm{o}},
\end{equation}
where
\begin{equation}\label{f_ou}
	\rho_{\tilde u}^{\mathrm{o}}\defeq\frac1{\Lambda}\sum_{\omega=1}^\nu \left(2\mu_\mathrm{o}-\mu_\omega\right)\overline W\left(\mathbf x_\omega-\tilde{\mathbf x}_u;\mathbf h_\omega,\boldsymbol{\uplambda}\right),
\end{equation}
where mirror bin $\tilde u$ is defined in analogy with \eqref{xu_mirr}. Expression \eqref{Dmirror_disc} is somewhat similar to \eqref{Nmirror_disc}; however now the simple reflection $\rho_{\tilde u}$ is replaced by a new term, $\rho_{\tilde u}^\mathrm{o}$. Examining expression \eqref{f_ou} we see that the only modification in the smoothing process is that, when transferring density from any non-empty bin $\omega$ to an external bin $\tilde u$, it must be done as if the bin $\omega$ contained a virtual number of particles equal to $2\mu_\mathrm{o}-\mu_\mathrm{\omega}$, instead of the actual value of $\mu_\omega$. 




\subsubsection{Robin}\label{subsub:Robin}
If we assume in \eqref{Inlet_balance} that the outer side of the boundary is a reservoir with a prescribed concentration, i.e., $\left(\phi\mathbf{D}\boldsymbol \nabla c\right)^-=0$ and $c^-=c_\mathrm{o}$, what we obtain is a Robin (or third-type) boundary condition, which can be written for boundary $\mathfrak{R}$ as
\begin{equation}\label{Fourier_Inlet}
	\mathbf{n}^\mathrm{T}\left(\mathbf{x}_\mathfrak{R}\right)\mathbf{D}\left(\mathbf{x}_\mathfrak{R}\right)\boldsymbol{\nabla}c\left(\mathbf{x}_\mathfrak{R}\right)=\mathbf{n}^\mathrm{T}\left(\mathbf{x}_\mathfrak{R}\right)\mathbf{v}\left(\mathbf{x}_\mathfrak{R}\right)\left(c\left(\mathbf{x}_\mathfrak{R}\right)-c_\mathrm{o}\right),\qquad \mathbf{x}_\mathfrak{R}\in \mathfrak{R}
\end{equation}
where $\mathbf{v}=\mathbf{q}/\phi$. Like in the previous cases, we treat the kernel as a very fast ``diffusive'' process, which interacts with the boundary analogously to the Green's function of the simulated dispersion. Since, for very large $\mathbf{D}$, condition \eqref{Fourier_Inlet} converges to \eqref{Neumann}, a Robin boundary affects the kernel identically to a homogeneous Neumann boundary (see $\S$\ref{sub:noflux}):
\begin{equation}\label{Rmirror_disc}
	\rho_{\mathfrak{R},u}=\rho_u+\rho_{\tilde u}.
\end{equation}
This would also apply to similar Robin boundary conditions originating from other assumptions, such as reactive walls \cite[e.g.,][]{Boccardo_Robin_18}.

\subsection{Extension to Irregular Boundaries}\label{sub:irreg}

The reflection principles given in $\S$\ref{sub:regular} are derived based on two assumptions: that the boundaries are regular and that they are aligned with the principal directions of the kernel. The latter can always be fulfilled by using an isotropic kernel (restricting the degrees of freedom of the bandwidth, see \cite{Sole2018}). However, the assumption of regular boundaries may just not be fulfilled, and then expressions \eqref{Nmirror_disc}, \eqref{Dmirror_disc} and \eqref{Rmirror_disc} are not valid as they rely on substituting the mirroring of the kernel for that of the evaluation point through \eqref{interchangeable}. Moreover, in that case, the true Green's function of the diffusion process associated to bandwidth $\mathbf{h}_\omega$ is much more complex than the solutions given by \eqref{W_N}, \eqref{absorbing} and \eqref{justDbound_W}. Nevertheless, as an approximation, one can use these solutions to modify the kernel directly, defining the mirror points by reflection through the closest boundary point, and with a weighting parameter to compensate for the irregularity. For impermeable or Robin, we propose the following kernel:
\begin{equation}\label{W_N_irreg}
	\overline W_\mathfrak{N}\left(\mathbf{x}_\omega,\mathbf{x}_u;\mathbf{h}_\omega,\boldsymbol{\uplambda}\right)\defeq \overline W\left(\mathbf{x}_\omega-\mathbf{x}_u;\mathbf{h}_\omega,\boldsymbol{\uplambda}\right)+\eta_\omega\overline W\left(\tilde{\mathbf{x}}_\omega-{\mathbf{x}}_u;\mathbf{h}_\omega,\boldsymbol{\uplambda}\right),
\end{equation}
where $\eta_{\omega}$ is a weighting parameter such that the mass of the ``reflection'' kernel $\overline W\left(\tilde{\mathbf{x}}_\omega-{\mathbf{x}}_u;\mathbf{h}_\omega,\boldsymbol{\uplambda}\right)$ entering the domain matches the mass of the ``regular'' kernel $\overline W\left(\mathbf{x}_\omega-\mathbf{x}_u;\mathbf{h}_\omega,\boldsymbol{\uplambda}\right)$ falling outside the domain. Thus, $\eta_\omega$ will be higher than one for a convex boundary and lower than one for a concave boundary. Essentially, the problem of non-bijection of the mirror points is fixed through this parameter. For Dirichlet boundary conditions:
\begin{equation}\label{W_D_irreg}
\overline W_\mathfrak{D}\left(\mathbf{x}_\omega,\mathbf{x}_u;\mathbf{h}_\omega,\boldsymbol{\uplambda}\right)\defeq \overline W\left(\mathbf{x}_\omega-\mathbf{x}_u;\mathbf{h}_\omega,\boldsymbol{\uplambda}\right)+\eta_\omega\left(\frac{2\mu_\mathrm{o}}{\mu_\omega}-1\right)\overline W\left(\tilde{\mathbf{x}}_\omega-{\mathbf{x}}_u;\mathbf{h}_\omega,\boldsymbol{\uplambda}\right),
\end{equation}
with the same definition for $\eta_\omega$. In the event that, because of the irregularity, the use of kernel \eqref{W_D_irreg} may generate negative density estimations in some bins, they should be set to zero. 

See $\S$\ref{subsub:irreg_comp} for an example implementation of the proposed approach.

\section{Computational Investigations and Discussion}\label{sec:comput}

In this section we investigate the performance of the proposed methodology on hypothetical RWPT models, with a focus on the features that are new with respect to the original approach introduced in \cite{Sole2018}. This includes the Eulerian grid projection, the iterative nature of the local kernel optimization, and the accounting for boundary conditions, both regular and irregular. In \ref{app:gsigma} we also discuss the use of local auxiliary parameters in the optimization process. We study each of these aspects separately, and then we focus on fixed-time concentration estimations rather than full reactive simulations, as it should be clear that a better concentration estimation will result in improved reactive transport modeling \cite{Sole2018}.

\subsection{Grid-Projected KDE}\label{sub:effic}

\begin{figure}[t]%
	\includegraphics[width=1\textwidth]{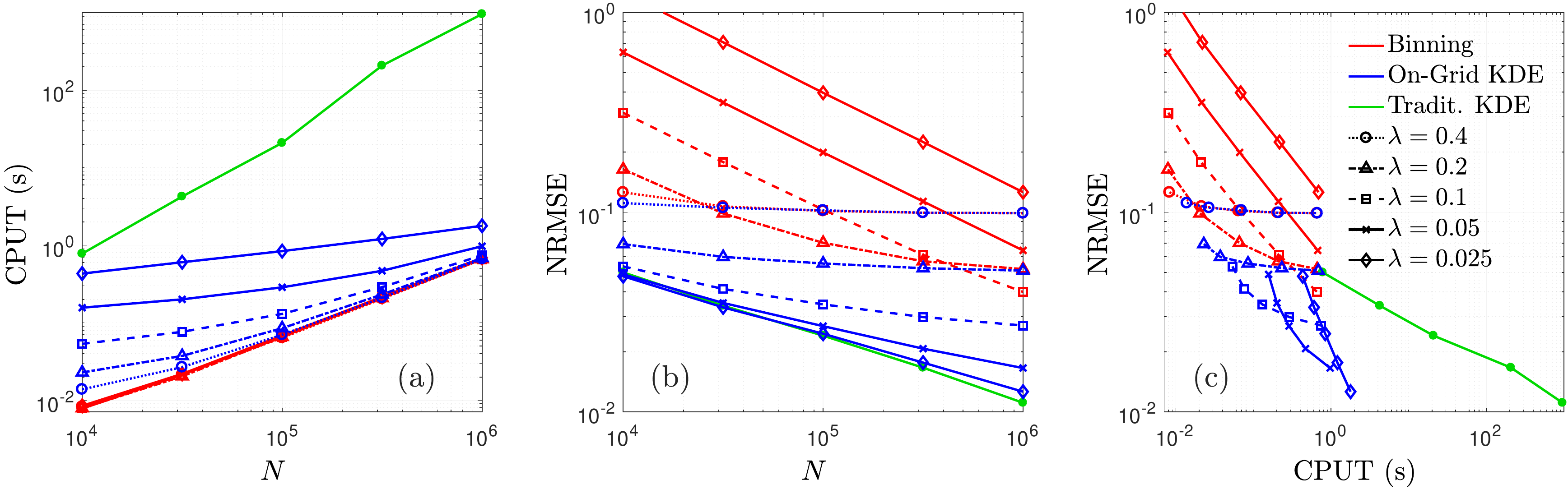}
	\centering
	\caption{For a density estimation given the setup described in $\S$\ref{sub:effic}, using different methodologies and bin sizes: (a) CPU Time invested as a function of the number of particles, (b) Normalized Root Mean Squared Error (NRMSE) as a function of the number of particles, (c) Resulting relationship between CPUT and NRMSE.}
	\label{fig:err}
\end{figure}

The use of kernel functions in particle-based methodologies is common for non-linear processes that involve interaction between individual particles, such as chemical reactions. Traditionally, the representation of the support volume of a particle as a smoothing kernel results in a loop through all pairs of potentially interacting numerical particles \cite{Sole-Mari2017}. Let us consider the simple example of an irreversible bimolecular reaction of type,
\begin{equation}\label{react}
\mathrm{A}+\mathrm{B}\rightarrow\mathrm{C},
\end{equation}
where $\mathrm{A}$, $\mathrm{B}$ and $\mathrm{C}$ are chemical compounds, and the ``batch'' reaction kinetics (that is, neglecting transport) can be described as:
\begin{equation}\label{kinetics}
\frac{\partial c_\mathrm{C}}{\partial t}=-\frac{\partial c_\mathrm{A}}{\partial t}=-\frac{\partial c_\mathrm{B}}{\partial t}=kc_\mathrm{A}c_\mathrm{B},
\end{equation}
where $k$ is the kinetic reaction constant and $c_\mathrm{A}$, $c_\mathrm{B}$ and $c_\mathrm{C}$ are the concentrations of compounds $\mathrm{A}$,  $\mathrm{B}$ and $\mathrm{C}$, respectively. In this case, for a particle $a$ of compound $\mathrm{A}$, we could estimate its probability of reaction in a time step $\Delta t$ as \cite{Sole2018}:
\begin{equation}\label{probreac}
P=k\Delta t c_\mathrm{B}(\mathbf{X}_a)=\frac{1}{\phi}k\Delta t m_\mathrm{B}\rho_\mathrm{B}(\mathbf{X}_a).
\end{equation}
For the sake of simplicity, $\phi$ in \eqref{probreac} is assumed constant. Here, the density of B-particles $\rho_\mathrm{B}$ is estimated from equations \eqref{c} and \eqref{density} at position $\mathbf{X}_a$ of the A-particle $a$. Doing this for all potentially reactive $\mathrm{A}$-particles would involve a double loop to see interaction between all particle pairs, and therefore scales in number of calculations as $N_\mathrm{A}N_\mathrm{B}$. In addition, it also requires a search algorithm, which would scale at best as $N_\mathrm{A}\log N_\mathrm{B}$. Aside from incorporating several kinds of boundary conditions, as addressed in $\S$\ref{sec:bounds}, a powerful argument for performing a pilot binning before the kernel smoothing is to increase computational efficiency, by pre-grouping the particles and avoiding the need for search algorithms; additionally one can pre-compute and store the matrix kernels, thus not having to evaluate the kernel function continuously. In this way, we simultaneously benefit from the low computational demands of binning as well as the accuracy of KDE. 

In order to exemplify this, consider a simple hypothetical 2D problem where, at a given time, compounds $\mathrm{A}$ and $\mathrm{B}$, each represented by $N_\mathrm{A}=N_\mathrm{B}=N$ particles, are distributed in space as partially overlapping multi-Gaussian distributions. Both these distributions have isotropic, unit variance $\sigma^2=1$, and their centers $\langle\mathbf{X}_\mathrm{A}\rangle$, $\langle\mathbf{X}_\mathrm{B}\rangle$, are separated by a distance $0.8\sigma$.

To perform a reactive time step given the described conditions, we consider three possible alternatives to estimate $\rho_\mathrm{B}(\mathbf{X}_a)$ in \eqref{probreac}, for all $a=1,\dots,N$: (i) Binning, (ii) Traditional KDE (eq. \eqref{density}), and (iii) The KDE method proposed herein (eq. \eqref{f_disc}). 

To independently evaluate and compare these techniques, we use a constant (global), isotropic kernel bandwidth of size $\hat h=\sigma N^{-\frac16}$ (see equation \eqref{h}). In the limit of $N\rightarrow\infty$ and $\boldsymbol{\uplambda}\rightarrow\mathbf{0}$, the estimated density at a point should converge to the true solution 
\begin{equation}\label{truedens}
\rho_\mathrm{B}^*(\mathbf{X}_a)=\frac{N}{2\pi\sigma^2}\exp{\left(-\frac{\left[\mathbf{X}_a-\langle\mathbf{X}_\mathrm{B}\rangle\right]^2}{2\sigma^2}\right)}.
\end{equation}
We measure the difference between $\rho_\mathrm{B}$ and $\rho_\mathrm{B}^*$ through the normalized root mean squared error (NRMSE):
\begin{equation}\label{NRMSE}
\mathrm{NRMSE}\defeq\left[\frac{\sum_{a=1}^{N}\left[\rho_\mathrm{B}(\mathbf{X}_a)-\rho_\mathrm{B}^*(\mathbf{X}_a)\right]^2}{\sum_{a=1}^{N}\left[\rho_\mathrm{B}^*(\mathbf{X}_a)\right]^2}\right]^\frac12.
\end{equation}
Additionally, we compute the time spent to perform the density estimation.  Figure \ref{fig:err} compares $\mathrm{NRMSE}$, $N$ and CPU time, given different values of isotropic bin size $\lambda$. 

For fixed $N$ and $\lambda$, the proposed density estimation technique is always more accurate than binning and more efficient than the traditional KDE. As a result, given a desired degree of accuracy, or equivalently, a fixed spatial scale of interest $\lambda$, we have strong evidence that the proposed technique is the optimal choice, in terms of attainable computational efficiency, for estimating  concentrations and reaction rates. For a very high value of $N$, as the kernel bandwidth $\hat h$ becomes small in comparison to $\lambda$, the proposed method and binning converge in terms of performance and accuracy, because there is in fact no effective smoothing. Prior to reaching that point (and mostly, in areas of lower particle density), the kernels are able to successfully and efficiently make up for the lack of particles.

\subsection{Local Bandwidth Optimization}\label{sub:conv}

\begin{figure}[t]%
	\centering
	\includegraphics[width=1\textwidth]{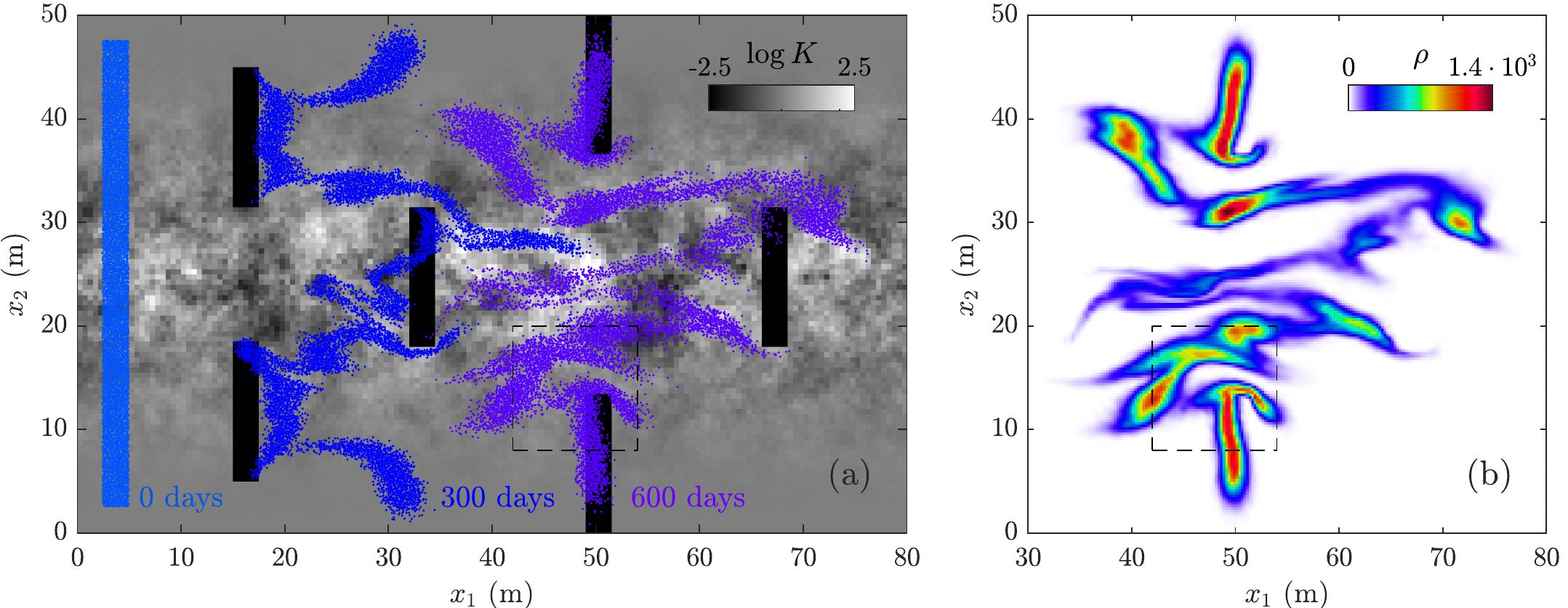}
	\caption{(a) Setup of the sample transport problem described in $\S$\ref{sub:conv}, and resulting particle distribution after 300 and 600 days. The dashed line signals the zoomed-in region of Figures \ref{fig:conv} and \ref{fig:robu}. (b) Particle densities, estimated by the novel method, at $t=600 \ \mathrm{days}$.}
	\label{fig:setup}
\end{figure}

\begin{figure}[t]%
	\centering
	\includegraphics[width=1\textwidth]{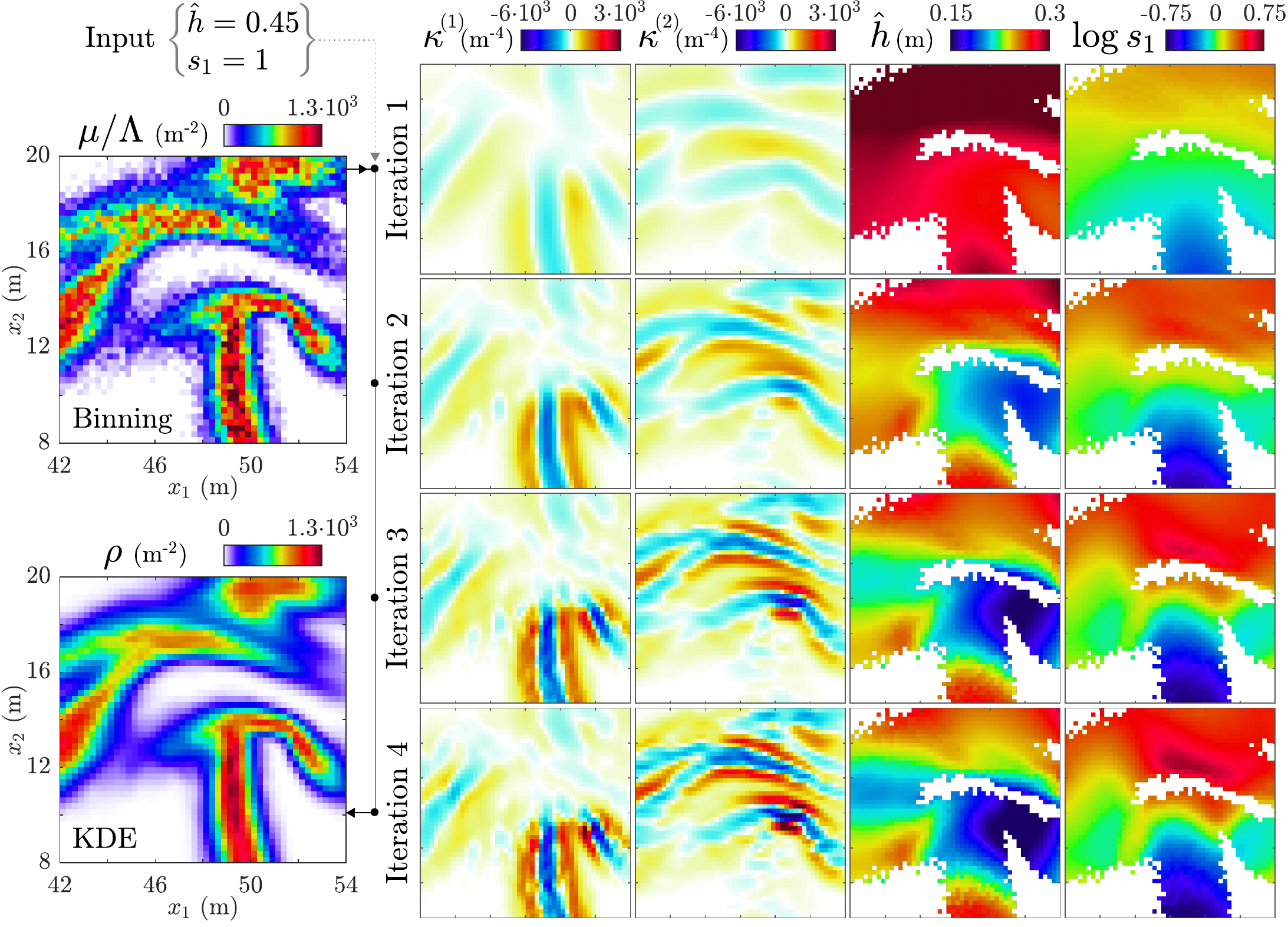}
	\caption{Graphical scheme of the kernel bandwidth optimization process, in the subregion marked by a dashed line in Figure \ref{fig:setup}, for an initial constant isotropic bandwidth of size $\hat h=0.45$. On the upper-left, the pilot binning density estimation. On the right, for each iteration, the estimated directional density curvatures, and the resulting size ($\hat h$) and rightwards elongation ($s_1$) of the kernel bandwidth. On the lower-left, the final kernel density estimation.}
	\label{fig:conv}
\end{figure}

Next, the local kernel bandwidth optimization and density estimation algorithm is tested in a synthetic example of advective-dispersive transport in a 2D heterogeneous porous medium (see Figure \ref{fig:setup}(a)). The spatial distribution of log hydraulic conductivity $K$, $Y=\log K$, in this domain of size ($80\times50 \ \mathrm{m}$) is built in 3 steps: first, (i) a random multi-Gaussian field  is generated with mean $\langle Y \rangle=0$, variance $\langle Y^2 \rangle=1$, and an exponential isotropic variogram with integral scale $I_Y=3 \ \mathrm{m}$; (ii) it is then multiplied with a field that evolves linearly in the vertical direction from 0 at the top and bottom boundaries to 1 at the center; (iii) finally, some low-conductivity ($Y=-2.5$) inclusions measuring $2.5\times13.5 \ \mathrm{m}$ are added as shown in Figure \ref{fig:setup}(a). The objective of this multistep procedure is to generate multiple different local features, that could be used as a test for the local bandwidth selection algorithm. 

Water flows through the porous medium following:
\begin{equation}\label{flow}
\nabla\cdot \mathbf{q}=0,\qquad\mathbf{q}=-K\nabla H,
\end{equation}
where $\mathbf{q}$ is the Darcy velocity and $H$ is the hydraulic head. The top and bottom boundaries are impermeable, and the head is prescribed at the left and right boundaries to force a mean hydraulic gradient of $2.5\%$, generating a mean flow from left to right.
The domain is discretized in square cells of $0.5\times0.5 \ \mathrm{m}$, and $\mathbf{q}$ at the cell interfaces is obtained via MODFLOW 2005 \cite{HarbaughArlen2005}. 

As an initial condition for transport, a uniform, rectangular distribution of $1.8\cdot10^5$ particles (representing a solute) is injected at $t=0$ near the left boundary, as shown in Figure \ref{fig:setup}. For time intervals $\left[t,t+\Delta t\right]$, particles move by random walk particle tracking (RWPT) \cite{Salamon2006a}:
\begin{equation}\label{rwpt}
\mathbf{X}_p\left(t+\Delta t\right)=\mathbf{X}_p\left(t\right)+\mathbf{A}\left(\mathbf{X}_p\left(t\right)\right)\Delta t+\mathbf{B}\left(\mathbf{X}_p\left(t\right)\right)\boldsymbol{\upxi}\sqrt{\Delta t},
\end{equation}
which is equivalent to solving the advection-dispersion equation (ADE). In \eqref{rwpt}, $\mathbf{A}\defeq\frac{1}{\phi}\left[\mathbf{q}+\nabla\cdot\left(\phi\mathbf{D}\right)\right]$, with $\mathbf{D}$ being the dispersion tensor; $\mathbf{B}$ is a $d\times d$ matrix such that $\mathbf{B}\mathbf{B}^{\mathrm{T}}=2\mathbf{D}$; and $\boldsymbol{\upxi}$ is a $d\times1$ vector of standard-normally distributed random numbers, uncorrelated in time. The spatially variable, anisotropic dispersion tensor $\mathbf{D}$ is determined following \cite{Bear2010}, with a longitudinal dispersivity of $\alpha_\ell=2\cdot10^{-3} \ \mathrm{m}$, a transverse dispersivity of $\alpha_t=3\cdot10^{-4} \ \mathrm{m}$, and a molecular diffusion of $D_m=2\cdot10^{-4} \ \mathrm{m}^2/\mathrm{day}$.
The spatial interpolation of velocities and the dispersion tensor is done using the hybrid linear-bilinear method proposed by \cite{LaBolle1996}. 

Figure \ref{fig:setup}(b) shows the spatial distribution of the particle density $\rho$ after 600 simulated days, estimated from the particle position information using the methodology presented in $\S$\ref{sec:ongrid}. In Figure \ref{fig:conv}, we show the iterative process of bandwidth differentiation and convergence (see $\S$\ref{sub:optimization}), starting from an arbitrary, uniform isotropic bandwidth with $\hat h=0.45 \ \mathrm{m}$, within a zoomed-in region of the domain (delimited by the dashed lines shown in Figure \ref{fig:setup}). The estimation of the curvatures $\kappa^{(i)}$, included in Figure \ref{fig:conv}, is crucial for the correct determination of the locally optimal scale ($\hat h$) and elongation ($s_1$) of the smoothing kernel bandwidth. We see for this particular example that, after 4 iterations, the solution has nearly stabilized. The smoothing of the ``pilot'' (binning) concentrations through this optimal local kernel is able to visibly reduce the noise, without generating excessive over-smoothing (see the two plots on the left of Figure \ref{fig:conv}). As a result, the NRMSE was reduced from a 93.7\% to a 7.0\% after the smoothing, with the NRMSE in this case being defined as in \eqref{NRMSE}, but for $\rho$ on all $\mathbf{x}_u$, and with the ``true'' solution in this case being approximated as the binning solution obtained for $N=8.64\cdot10^7$ ($\sim 2.5$ orders of magnitude more than the test case).  

\begin{figure}[t]%
	\centering
	\includegraphics[width=1\textwidth]{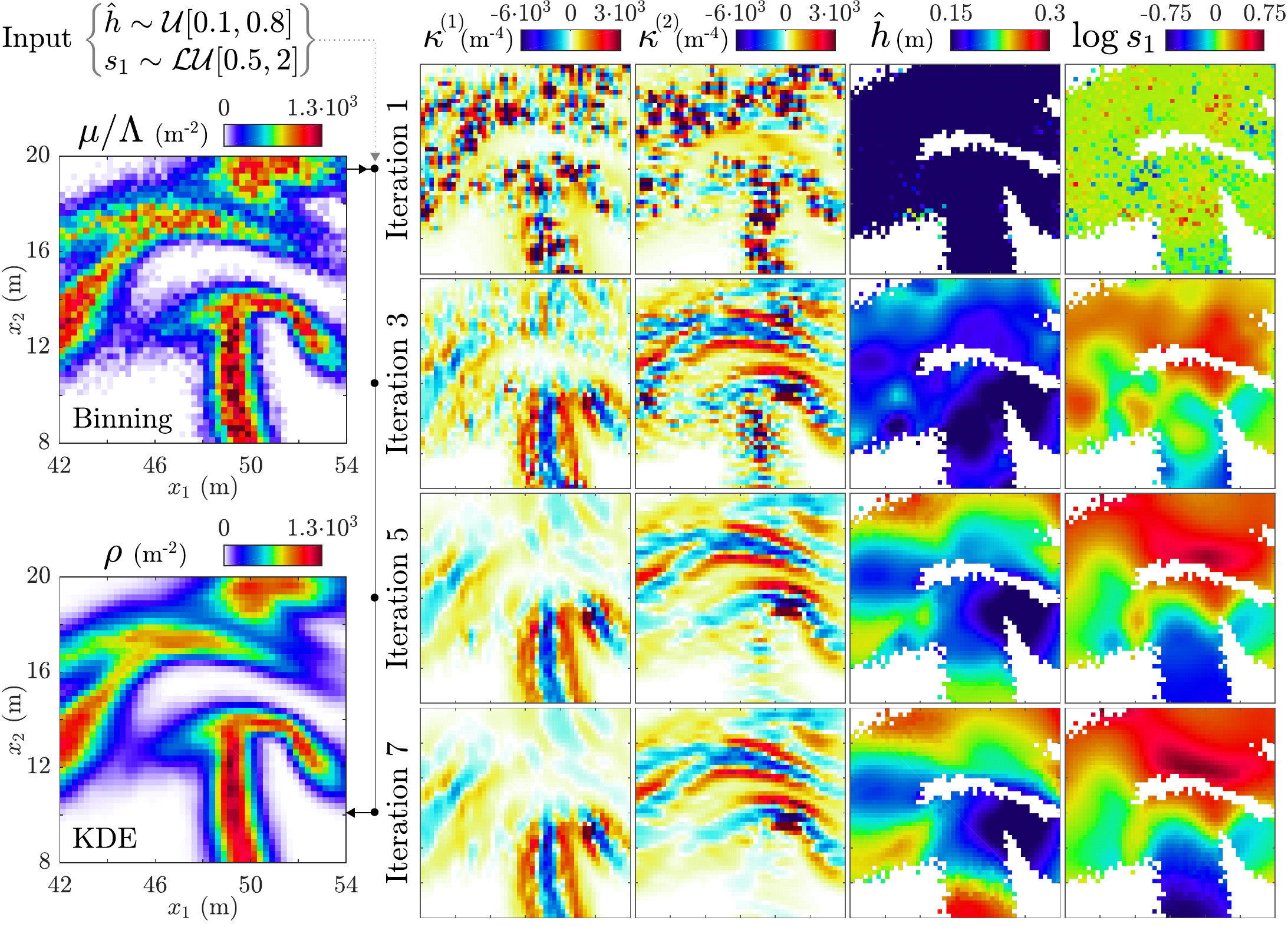}
	\caption{Graphical scheme of the kernel bandwidth optimization process, in the region marked by a dashed line in Figure \ref{fig:setup}, for an initial uncorrelated random bandwidth, with uniformly distributed size and log-uniformly distributed rightwards elongation. On the upper-left, the pilot binning density estimation. On the right, for each iteration (1,3,5,7), the estimated directional density curvatures, and the resulting size ($\hat h$) and rightwards elongation ($s_1$) of the kernel bandwidth. On the lower-left, the final kernel density estimation.}
	\label{fig:robu}
\end{figure}

Another relevant property of the iterative optimization algorithm is its robustness, i.e. its ability to reach the same solution given different initial values. This is particularly important in the context of particle tracking simulations, where the particles can ``carry'' the support volume for some number of steps and then use it as an input to update the next optimal support volume, conferring it an evolutionary nature. For this to make sense, it is critical that for any given current distribution of particles, the solution always converges to a unique set of values, regardless of the history of the local kernel bandwidth. 

To test this, we repeat the previous numerical experiment, but this time, the initial bandwidth is random and uncorrelated in space, with $\hat h$ drawn from a uniform distribution with lower and upper bounds of $0.1$ and $0.8$; and with $s_1$ drawn from a log-uniform distribution (the logarithm is uniformly distributed) with lower and upper limits of $0.5$ and $2.0$. These are deliberately chosen to be challenging for the convergence of the algorithm. As can be seen from Figure \ref{fig:robu}, in the first iteration this results in a curvature estimation ($\kappa^{(1)},\kappa^{(2)}$) that is far from the correct solution, and characterized by high absolute values and strong variations with no apparent spatial correlation. As a consequence, the first estimation of $\hat h$ and $s_1$ gives a small bandwidth without a well-defined elongation direction. Yet, after a few iterations, we see that the identification of the curvatures substantially improves, and concurrently the bandwidth scale and elongation start differentiating spatially distinct regions. For this highly adverse case of a locally random and uncorrelated input value, after only 7 iterations we have nearly reached the same stable solution as in Figure \ref{fig:conv}. 

\subsection{Boundary Conditions}\label{sub:bound_comp}

\begin{figure}[t]%
	\centering
	\includegraphics[width=0.75\textwidth]{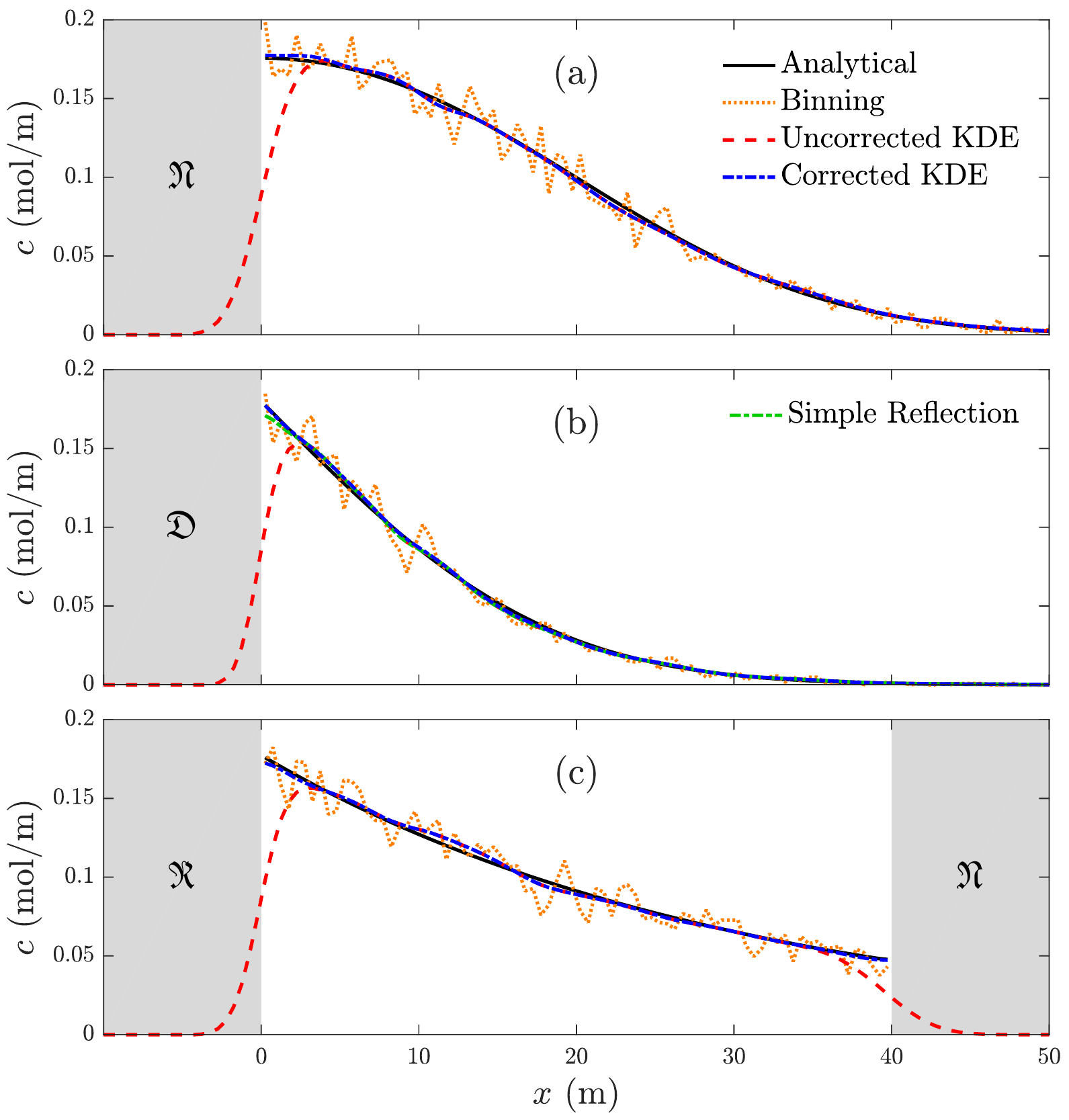}
	\caption{Comparison between the analytical and the RWPT solution, with different concentration estimation techniques, for the simple transport problems described in Section $\S$\ref{sub:bound_comp}. The spurious fluctuations in the pilot binning are corrected by the optimal kernel smoothing, at the cost of inaccuracies near the boundaries. This issue is solved by the boundary corrections introduced in $\S$\ref{sec:bounds}.}
	\label{fig:bounds1D}
\end{figure}

Here we implement and evaluate the boundary condition correction techniques described in $\S$\ref{sec:bounds}. First, we perform a concentration estimation near the boundary in three simple 1D RWPT settings. In all cases, the solute moves by dispersion with $D=0.1 \ \mathrm{m}^2/\mathrm{day}$ for a total time of $\tau=1000 \ \mathrm{days}$; the medium porosity is $\phi=0.25$. Each particle has a mass (or more precisely, an amount of substance) of $m=10^{-4} \ \mathrm{mol}$. The bin size is $\lambda=0.5 \ \mathrm{m}$. We compare the three estimation methods (the pilot binning, the uncorrected KDE, and the corrected KDE) to the analytical solution for each case.

\subsubsection{Diffusion near impermeable wall}

In the first example, the initial condition is a Dirac delta pulse of solute (with a total mass $M=1 \ \mathrm{mol}$) located at $x=10 \ \mathrm{m}$, near an impermeable boundary located at $x_\mathfrak{N}=0$. 
The results are shown in Figure \ref{fig:bounds1D}(a). Since here diffusion is the only physical process, and the corrected kernel emulates diffusion, the concentration estimation has an excellent agreement with the analytical solution. 

\subsubsection{Diffusion through boundary}

In the second example, all initial concentrations are zero inside the domain. There is a Dirichlet boundary condition such that $c_\mathrm{o}=0.18 \ \mathrm{mol}/\mathrm{m}$ at $x_\mathfrak{D}=0$. The results are shown in Figure \ref{fig:bounds1D}(b). The Dirichlet ``reflection'' technique \eqref{Dmirror_disc} is able to correctly reconstruct the concentration field and gradient near the boundary. It is worth noting that a simple reflection as in \eqref{Nmirror_disc} would have resulted in a zero-gradient instead, as shown by the curve labeled as ``Simple Reflection''. Nevertheless, the error involved in using \eqref{Nmirror_disc} instead of \eqref{Dmirror_disc}, at least in this specific case, was relatively small (compared, for instance, to the error associated with not performing a boundary correction at all).

\subsubsection{Column with inlet and outlet reservoirs}

In this third example we consider a constant rightwards advection ($q=0.055 \ \mathrm{m}/\mathrm{day}$) along with a linear degradation:

\begin{equation}\label{degr}
\frac{\partial c}{\partial t}=-\frac{q}{\phi}\frac{\partial c}{\partial x}+D\frac{\partial^2 c}{\partial x^2}-k c.
\end{equation}

The reaction is simulated stochastically by a particle reaction probability $P=k\Delta t$ (i.e., this is the probability a particle dies in any given time step). At the inlet we have a Robin boundary condition as described in $\S$\ref{subsub:Robin}, and at the outlet we set a zero-gradient (homogeneous Neumann) condition. This problem has a stationary solution. From Figure \ref{fig:bounds1D}(c), we see that the corrected density estimation has a zero-gradient near the boundaries, and that the gradient is not zero in the true solution. This is because the kernel reflection method is based on pure diffusion and unable to account for the gradient generated by the combined action of advection and reaction. Nonetheless, one can appreciate in Figure \ref{fig:bounds1D}(c) a substantial improvement for the corrected KDE method with respect to the uncorrected one. 

\subsubsection{Irregular Boundaries}\label{subsub:irreg_comp}

\begin{figure}[t]%
	\centering
	\includegraphics[width=0.75\textwidth]{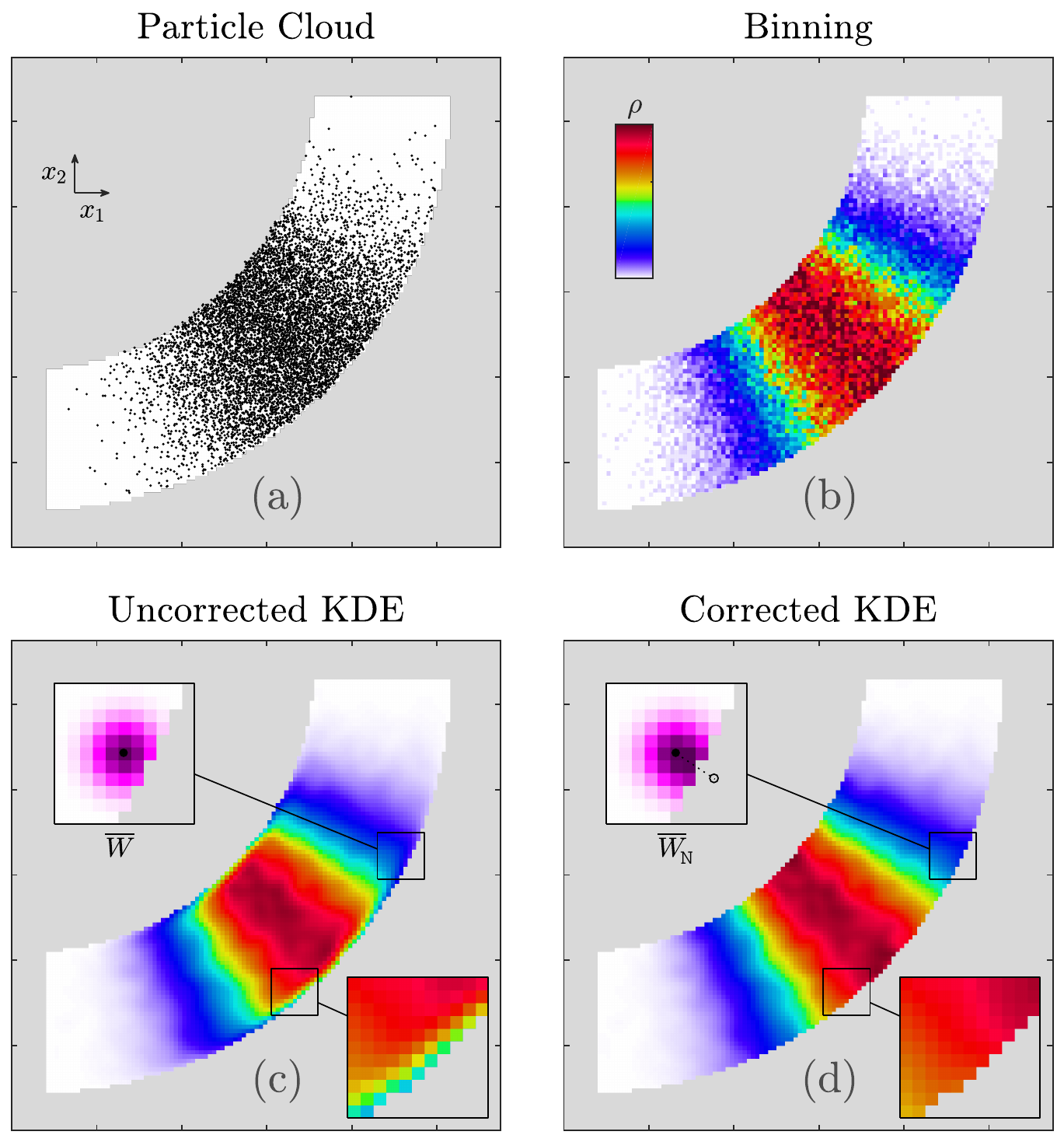}
	\caption{Graphical example of a kernel density estimation with irregular boundaries, for an impermeable boundary condition. The uncorrected KDE (c) produces an unphysical loss of mass near the boundary. This is corrected by a ``mirror'' modification of the kernels near the boundary (d).}
	\label{fig:irreg}
\end{figure}

In order to illustrate the boundary correction technique for the case of irregular boundaries explained in $\S$\ref{sub:irreg}, we designed a problem consisting of a domain with the shape of an arched tube (Figure \ref{fig:irreg}) with impermeable boundaries. The region shown has dimensions $115\times115$, and the spatial discretization is square with $\lambda=1$. The particle cloud (Figure \ref{fig:irreg}(a)) is the result of a random injection of $100\,000$ particles. In polar coordinates (the origin being the center of curvature of the tube), the square of the radial coordinate of the particle positions is uniformly-distributed, whereas the angular coordinate is Gaussian-distributed. Such a distribution could be thought of as a hypothetical result of advection-dispersion through the tube. 

Despite the high number of particles used, we observe in Figure \ref{fig:irreg}(b) a considerable amount of noise in the binning estimation. This issue is fixed by our locally adaptive kernel smoothing technique, as shown in Figure \ref{fig:irreg}(c), but at the expense (prior to any boundary correction) of an artificial density loss near the boundaries (see zoomed-in region). As illustrated by the representation of the kernel $\overline W$ near the boundary, this is caused by the lack of a reflection of the mass lost through the boundary by the smoothing. In Figure \ref{fig:irreg}(d), we see that the modified kernel (see zoomed-in representation) successfully fixes the loss of mass through the boundary, yielding a satisfactory reconstruction of the concentrations.

\section{Summary and Conclusions}\label{sec:conc}

We have presented a novel technique to estimate particle densities using the limited amount of information provided by a finite sample of particle positions. Although the spectrum of possible applications is wide, the focus of this work is the reconstruction of solute concentrations in random walk particle tracking (RWPT) simulations, which is relevant for the visualization of results and the incorporation of chemical reactions. Our technique relies on the accuracy of locally adaptive kernel density estimation (KDE), which is implemented in combination with a spatial discretization, resulting in benefits including computational efficiency and accurate implementation of boundary conditions. The method is valid in 1, 2 or 3 spatial dimensions. In principle, it can be applied to Lagrangian modeling techniques other than RWPT, or even to other applications of density estimation. An open-source MATLAB code, ``Bounded Adaptive Kernel Smoothing'' (BAKS), has been developed and made available to the community as a result of this work \cite{baks}. 

Our computational investigations provide strong evidence that our proposed methodology deals well with the dilemma between the accuracy of kernel methods and the low computational requirements of binning. From our tests, we see that the desired degree of accuracy (characterized by the choice of bin size $\boldsymbol{\uplambda}$) is always achieved faster (in terms of invested computational effort) with our proposed methodology compared with binning. The two methods converge, both in accuracy and performance, for high particle numbers, but prior to reaching that point, the kernel approach reaches an acceptable level of error earlier. 
Likewise, direct Lagrangian utilization of the kernel functions involves a worse scaling of CPU time with particle number than our method. As a result, the proposed method can achieve the same level of accuracy with lower computational effort.

Through a sample simulation of conservative transport in a heterogeneous porous medium, we show the convergence of the local kernel bandwidth iterative optimization method towards a stable result. Using that optimal local bandwidth to smooth the pilot binning density estimation resulted in a reduction of normalized error from $93.7\%$ to $7.0\%$. We also demonstrate the robustness of the method, in terms of being able to reach this same solution regardless of the initial input values provided. The fully local nature of the optimization process, which is a novel aspect with respect to the original methodology \cite{Sole2018}, allows the kernel to achieve a higher degree of local differentiation in terms of size and shape.

Near boundaries, the kernel is assumed to emulate the Green's function of a fast diffusion process. Hence, the kernel is affected by the boundary following the analytical solution of pure diffusion corresponding to the relevant boundary condition, which results in simple and efficient reflection rules. The case of irregular boundaries is slightly more complicated, as it requires individual modification of each kernel, with a weighting of the reflection to ensure proper mass conservation. The simple implementation examples illustrate the importance of including these boundary corrections in the density estimation.

\appendix
\section{Details on treatment of matrix kernels}\label{app:matrices}
The projected kernels $\overline W$ and $\overline V$, introduced in $\S$\ref{subsec:estimation} and $\S$\ref{sub:optimal_h}, respectively, require some corrections to ensure that they keep the main properties of the original kernels $W$, $V$, after the on-grid projection. In the case of $\overline W$, the original Gaussian kernel $W$ integrates exactly to 1 only over $\mathbb R^d$. If a cutoff distance is imposed, a normalization is needed to impose mass conservation:
\begin{equation}\label{normW}
\overline W^\prime\left(\boldsymbol{\uplambda}\odot\mathbf{z};\mathbf h,\boldsymbol{\uplambda}\right)\defeq\frac{\overline W\left(\boldsymbol{\uplambda}\odot\mathbf{z};\mathbf h,\boldsymbol{\uplambda}\right)}{\sum_{\pmb{\zeta}} \overline W\left(\boldsymbol{\uplambda}\odot\pmb{\zeta};\mathbf h,\boldsymbol{\uplambda}\right)},
\end{equation}
where $\overline W\left(\boldsymbol{\uplambda}\odot\mathbf{z};\mathbf h,\boldsymbol{\uplambda}\right)$ is the unmodified kernel at cell index $\mathbf{z}$ (equation \eqref{Wprime_lam}), $\mathbf{z}=\mathbf{0}$ being the cell where the center of the kernel is located, and $\overline W^\prime$ is the modified kernel. In \eqref{normW}, the denominator is a sum over all cells within the cutoff limits.

In the case of $\overline V$, there are two corrections that need to be performed in order to conserve the original purpose of $V$. On one hand, the positive values must be weighted so that the kernel integrates to zero within the cutoff limits:
\begin{equation}\label{V_prescaling}
\overline V^{\prime(i)}\left(\boldsymbol{\uplambda}\odot\mathbf{z};\mathbf g,\boldsymbol{\uplambda}\right)\defeq a\overline V^{(i)}\left(\boldsymbol{\uplambda}\odot\mathbf{z};\mathbf g,\boldsymbol{\uplambda}\right),
\end{equation}
\begin{equation}
a\defeq
\begin{cases}
-\frac{\sum_{\overline V^{(i)}<0}\overline V^{(i)}\left(\boldsymbol{\uplambda}\odot\pmb{\zeta};\mathbf{g},\boldsymbol{\uplambda}\right)}{\sum_{\overline V^{(i)}>0}\overline V^{(i)}\left(\boldsymbol{\uplambda}\odot\pmb{\zeta};\mathbf{g},\boldsymbol{\uplambda}\right)},& \text{if} \ \overline V^{(i)}>0 \\
\hfil 1, & \text{if} \ \overline V^{(i)}\leq 0
\end{cases}.
\end{equation}
A simple intuitive example of the importance of this correction is that a constant particle density must necessarily yield a zero-curvature estimation, and this will only occur if the curvature kernel has zero-mean. 

Besides, the final purpose of $\overline V^{(i)}$ is the estimation of the squared second spatial derivatives of the particle density ($\kappa^{(i)}_\omega\kappa^{(j)}_\omega$ in \eqref{psi_disc}). The spatial averaging involved in the grid projection could lead to a systematic under-prediction of these values. To prevent that, we scale the kernel so that it keeps the $\mathrm{L}^2$-norm of the original kernel $V$ after the projection: 
\begin{equation}\label{V_scaling}
\overline V^{\prime\prime(i)}\left(\boldsymbol{\uplambda}\odot\mathbf{z};\mathbf{g},\boldsymbol{\uplambda}\right)\defeq\left[\frac{\Lambda\norm{V^{(i)}}^2}{\sum_{\pmb{\zeta}}\left[\overline V^{\prime(i)}\left(\boldsymbol{\uplambda}\odot\pmb{\zeta};\mathbf{g},\boldsymbol{\uplambda}\right)\right]^2}\right]^\frac12\cdot\overline V^{\prime(i)}\left(\boldsymbol{\uplambda}\odot\mathbf{z};\mathbf{g},\boldsymbol{\uplambda}\right),
\end{equation}
where $\norm{ \ }^2$ is the $\mathrm{L}^2$-norm operator, i.e.,
\begin{equation}\label{L2norms_V}
\norm{V^{(i)}}^2\defeq\int_{\mathbb{R}^d}\left[V^{(i)}\left(\mathbf{r};\mathbf{g}\right)\right]^2\mathrm d\mathbf r=\frac3{2^{(d+2)}\pi^{\frac d2}g_i^5\left(\prod_{j\neq i}g_j\right)}.
\end{equation}
$\overline V^{\prime\prime(i)}$ are the final values of the projected curvature kernel.

As mentioned in the main body of this work, the values of $\overline W$ and $\lambda_i^2\overline V^{(i)}$ only depend on the directional ratios between the bandwidth ($\mathbf{h}$ or $\mathbf{g}$) and the grid size ($\boldsymbol{\uplambda}$). Discretizing the values that these ratios are allowed to adopt, and imposing a cutoff distance (as in Figure \ref{fig:kernelmatrices}), $\overline W$ and $\lambda_i^2\overline V^{(i)}$ become finite sets of matrices with a finite number of entries. Repeated use of the same (or very similar) kernel bandwidth results in the exact same matrix kernel, which can be stored in the memory after its first generation and correction. Then, computation of \eqref{f_disc}, \eqref{kappa_disc}, \eqref{n_disc} and \eqref{psi_disc} only requires accessing the pre-computed matrix entries in the memory and performing the relevant weighted summation, avoiding redundant computational efforts.

\section{Derivation of expression \eqref{sigma_rel_2}}\label{app:22}

Within a virtual Gaussian distribution of $N_u^\sigma$ particles, centered at $\boldsymbol{\upmu}_u\equiv[\mu_{u,1},\dots,\mu_{u,d}]^\mathrm{T}$ with the vector of directional standard deviations $\boldsymbol{\upsigma}_u\equiv[\sigma_{u,1},\dots,\sigma_{u,d}]^\mathrm{T}$, 
\begin{equation}\label{rho_virt}
	\rho(\mathbf{x})=N_u^\sigma\prod_{i=1}^{d}\frac{1}{\sqrt{2\pi}\sigma_{u,i}}\exp{\left(-\frac{(x_i-\mu_{u,i})^2}{2\sigma_{u,i}^2}\right)}\equiv N_u^\sigma W\left(\mathbf{x}-\boldsymbol{\upmu}_u;\boldsymbol{\upsigma}_u\right).
\end{equation}
If $\boldsymbol{\upsigma}_u$ is also the bandwidth of the integration kernel, then following \eqref{n_disc},
\begin{equation}\label{n_virt}
\begin{split}
	n_u&=\int \rho\left(\mathbf x\right)W\left(\mathbf{x}-\mathbf{x}_u;\boldsymbol{\upsigma}_u\right)\mathrm d\mathbf x \\
	&=N_u^\sigma\int W\left(\mathbf{x}-\boldsymbol{\upmu}_u;\boldsymbol{\upsigma}_u\right)W\left(\mathbf{x}-\mathbf{x}_u;\boldsymbol{\upsigma}_u\right)\mathrm d\mathbf x \\
	&=N_u^\sigma W\left(\mathbf{x}_u-\boldsymbol{\upmu}_u;\sqrt{2}\boldsymbol{\upsigma}_u\right)=\frac{N_u^\sigma[{W\left(\mathbf{x}_u-\boldsymbol{\upmu}_u;\boldsymbol{\upsigma}_u\right)}]^{1/2}}{(8\pi\hat\sigma_u^2)^{d/4}} \\
	&=\sqrt{\frac{N_u^\sigma\rho(\mathbf{x}_u)}{(\sqrt{8\pi}\hat\sigma_u)^{d}}}\approx\sqrt{\frac{N_u^\sigma\rho_u}{(\sqrt{8\pi}\hat\sigma_u)^{d}}}.
\end{split}
\end{equation}
Here, the approximation $\rho(\mathbf{x}_u)\approx\rho_u$ is equivalent to the one in \eqref{n_disc}. Taking the square on both sides of expression \eqref{n_virt} and rearranging we obtain
\begin{equation}\label{sigma_rel_2_rep}
N_u^\sigma=\frac{\left(\sqrt{8\pi}\hat\sigma_u\right)^d n_u^2}{\rho_u},
\end{equation}
which is the expression given in \eqref{sigma_rel_2}.

\section{Curvature kernel bandwidth for Gaussian distribution}\label{app:gopt}

Following the approach of \cite[Appendix E]{botev2010}, here generalized for $d$ dimensions, the optimal isotropic $\hat g^{(i)}$, given a distribution of particles $\rho(\mathbf{x})$, will be
\begin{equation}\label{gopt_bot}
	\hat g^{(i)}=\left(\frac{(2+2^{-\frac{d}{2}-1})N}{(2\pi)^{\frac{d}{2}}\sum_{j=1}^{d}\int\left(\frac{\partial^3\rho}{\partial x_i^2\partial x_j}\right)^2\mathrm{d}\mathbf{x}}\right)^{\frac{1}{d+6}},
\end{equation}
with $N$ being the total number of particles. Taking $\rho(\mathbf{x})$ to be a Gaussian distribution of $N$ particles with a vector of directional standard deviations such that $\boldsymbol{\upsigma}=\hat \sigma\mathbf{s}$, with $\prod_{i=1}^ds_i=1$, we have
\begin{equation}\label{int}
	\int\left(\frac{\partial^3\rho}{\partial x_i^2\partial x_j}\right)^2\mathrm{d}\mathbf{x}=\frac{3(1+4\delta_{ij})N^2}{8(4\pi)^{\frac{d}{2}}\hat\sigma^{d+6}s_{i}^4s_{j}^2},
\end{equation}
and then substituting \eqref{int} in \eqref{gopt_bot} and rearranging we obtain
\begin{equation}\label{gopt_semirep}
\hat g^{(i)}=\left[\frac{4+2^{\frac{d}{2}+4}}{3N\sum_{j=1}^{d}\frac{1+4\delta_{ij}}{s_i^4s_j^2}}\right]^{\frac1{d+6}}\hat\sigma,
\end{equation}
which is equivalent to the combination of equations \eqref{gopt} and \eqref{theta} given in $\S$\ref{sub:g}.

\section{Local vs Global selection of $\boldsymbol{\upsigma}$ and $\mathbf{g}$}\label{app:gsigma}

\begin{figure}[t]%
	\centering
	\includegraphics[width=0.75\textwidth]{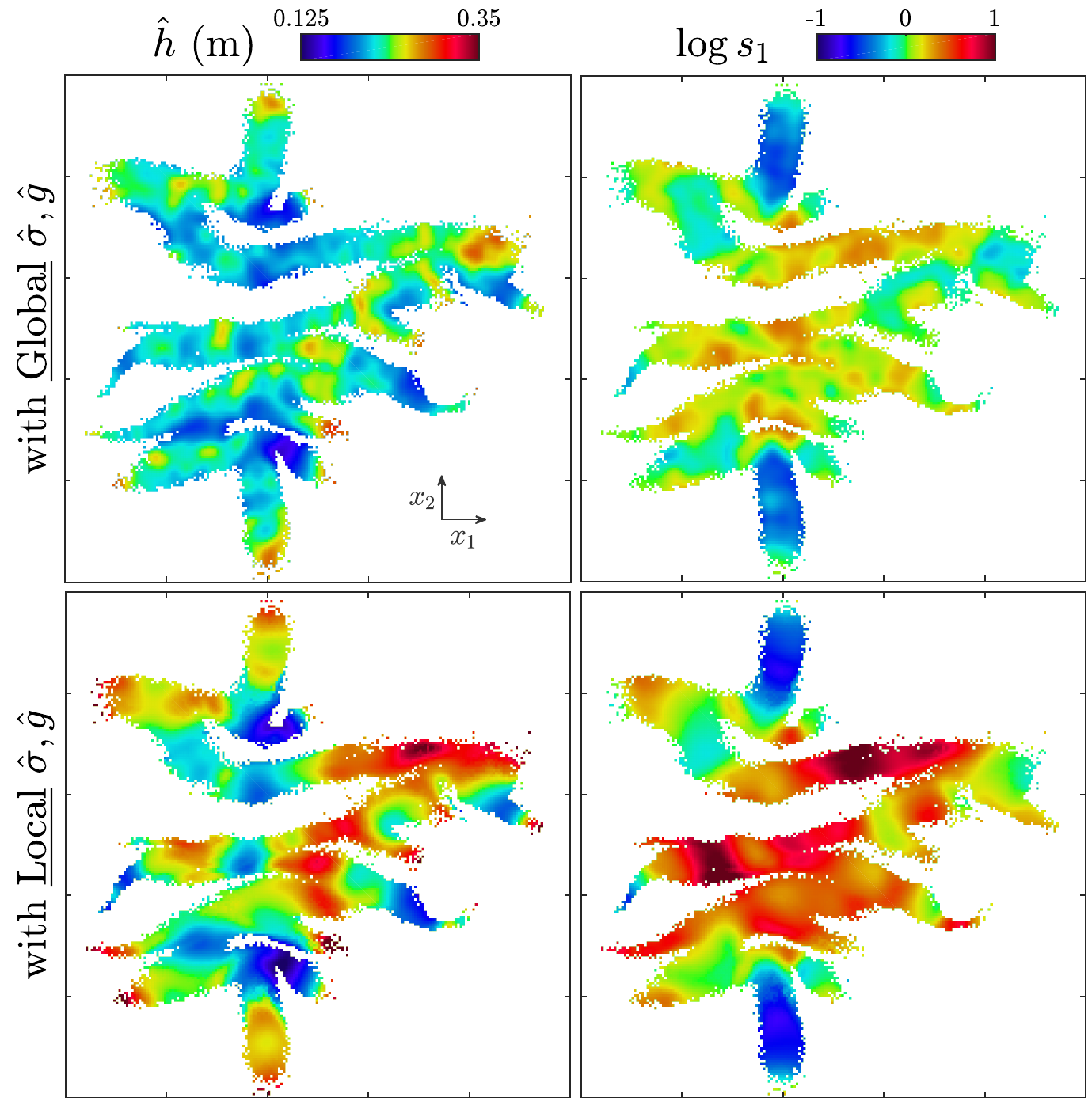}
	\caption{Graphical comparison of the size $\hat h$ (first column) and rightwards elongation $s_1$ (second column) of the kernel bandwidth obtained by optimization using a global integration kernel $\hat \sigma$ and curvature kernel $\hat g$ (first row), or local values instead (second row). Note the higher degree of local differentiation in the second case.}
	\label{fig:gsigma}
\end{figure}

Within the procedure presented in $\S$\ref{sec:ongrid}, the selection of the curvature kernel bandwidth ($\mathbf{g}_u^{(i)}$) at a bin $u$ is completely local and independent of all particles located outside the range determined by the integration support $\boldsymbol{\upsigma}_u$. In the original methodology presented in \cite{Sole2018}, a global $\mathbf{g}^{(i)}$ was 
used over the whole domain, as noted in $\S$\ref{sub:g}. As a result, the local kernel bandwidth would be indirectly conditioned by global features. For instance, if the plume was dominantly formed by strongly elongated shapes in one specific direction, the curvature estimation anywhere would be biased towards detecting those kinds of features. This would limit the ability of the kernel bandwidth to optimally adapt strictly to the nearby particle distribution. To illustrate this, we compare the distribution of local bandwidths $\mathbf{h}$ obtained with the method presented in $\S$\ref{sec:ongrid} to the one that we obtain when using global (instead of local) values for $\boldsymbol{\upsigma}$ and $\mathbf{g}$, following what is described in \cite{Sole2018}. We use the same example particle distribution of $\S$\ref{sub:conv}.

Figure \ref{fig:gsigma} shows the local bandwidth scale ($\hat h$) and elongation ($s_1$) values obtained with the two described approaches. We clearly observe a higher degree of differentiation, that is, an increased ability of the bandwidth scale and elongation to reach extreme values when using the novel, fully local methodology presented  here. This is particularly true for large bandwidths, which can only exist in the absence of noise in the curvature estimation (see Figure \ref{fig:robu} as an example). The distribution of bandwidth scales and elongations also appears to be significantly smoother (less affected by ``spurious'' fluctuations) with the novel methodology, which suggests a more accurate identification of the dominant shapes of the local particle distribution. 

Quantitatively, the NRMSE of $\rho$ changes from 7.4\% for the global parameter choice to the aforementioned 7.0\% when using the novel fully local methodology. Although this may seem like a small reduction, the difference would probably become larger in the case of an even stronger spatial differentiation of the local particle distributions.

\bibliography{BIB}

\begin{thebibliography}{10}
\expandafter\ifx\csname url\endcsname\relax
  \def\url#1{\texttt{#1}}\fi
\expandafter\ifx\csname urlprefix\endcsname\relax\def\urlprefix{URL }\fi
\expandafter\ifx\csname href\endcsname\relax
  \def\href#1#2{#2} \def\path#1{#1}\fi

\bibitem{Salamon2006a}
P.~Salamon, D.~Fern{\`{a}}ndez-Garcia, J.~J. G{\'{o}}mez-Hern{\'{a}}ndez, {A
  review and numerical assessment of the random walk particle tracking method},
  Journal of Contaminant Hydrology 87~(3-4) (2006) 277--305.
\newblock \href {https://doi.org/10.1016/j.jconhyd.2006.05.005}
  {\path{doi:10.1016/j.jconhyd.2006.05.005}}.

\bibitem{Salamon2006}
P.~Salamon, D.~Fern{\`{a}}ndez-Garcia, J.~J. G{\'{o}}mez-Hern{\'{a}}ndez,
  {Modeling mass transfer processes using random walk particle tracking}, Water
  Resources Research 42~(11).
\newblock \href {https://doi.org/10.1029/2006WR004927}
  {\path{doi:10.1029/2006WR004927}}.

\bibitem{Berkowitz2006}
B.~Berkowitz, A.~Cortis, M.~Dentz, H.~Scher, {Modeling Non-fickian transport in
  geological formations as a continuous time random walk}, Reviews of
  Geophysics 44~(2).
\newblock \href {https://doi.org/10.1029/2005RG000178}
  {\path{doi:10.1029/2005RG000178}}.

\bibitem{Benson2017}
D.~A. Benson, T.~Aquino, D.~Bolster, N.~Engdahl, C.~V. Henri,
  D.~Fern{\`{a}}ndez-Garcia, {A comparison of Eulerian and Lagrangian transport
  and non-linear reaction algorithms}, Advances in Water Resources 99 (2017)
  15--37.
\newblock \href {https://doi.org/10.1016/j.advwatres.2016.11.003}
  {\path{doi:10.1016/j.advwatres.2016.11.003}}.

\bibitem{engdahl_ddc}
N.~B. Engdahl, M.~J. Schmidt, D.~A. Benson, Accelerating and parallelizing
  lagrangian simulations of mixing-limited reactive transport, Water Resources
  Research 55.
\newblock \href {https://doi.org/10.1029/2018WR024361}
  {\path{doi:10.1029/2018WR024361}}.

\bibitem{Benson2008}
D.~A. Benson, M.~M. Meerschaert, {Simulation of chemical reaction via particle
  tracking: Diffusion-limited versus thermodynamic rate-limited regimes}, Water
  Resources Research 44~(12).
\newblock \href {https://doi.org/10.1029/2008WR007111}
  {\path{doi:10.1029/2008WR007111}}.

\bibitem{Paster2013}
A.~Paster, D.~Bolster, D.~A. Benson, {Particle tracking and the
  diffusion-reaction equation}, Water Resources Research 49~(1) (2013) 1--6.
\newblock \href {https://doi.org/10.1029/2012WR012444}
  {\path{doi:10.1029/2012WR012444}}.

\bibitem{Bolster2016}
D.~Bolster, A.~Paster, D.~A. Benson, {A particle number conserving Lagrangian
  method for mixing-driven reactive transport}, Water Resources Research 52~(2)
  (2016) 1518--1527.
\newblock \href {https://doi.org/10.1002/2015WR018310}
  {\path{doi:10.1002/2015WR018310}}.

\bibitem{Lazaro2019}
L.~J. Perez, J.~J. Hidalgo, M.~Dentz, Upscaling of mixing-limited bimolecular
  chemical reactions in poiseuille flow, Water Resources Research 55~(1) (2019)
  249--269.
\newblock \href {https://doi.org/10.1029/2018WR022730}
  {\path{doi:10.1029/2018WR022730}}.

\bibitem{Paster2014}
A.~Paster, D.~Bolster, D.~A. Benson, {Connecting the dots: Semi-analytical and
  random walk numerical solutions of the diffusion-reaction equation with
  stochastic initial conditions}, Journal of Computational Physics 263 (2014)
  91--112.
\newblock \href {https://doi.org/10.1016/j.jcp.2014.01.020}
  {\path{doi:10.1016/j.jcp.2014.01.020}}.

\bibitem{Ding2017}
D.~Ding, D.~A. Benson, D.~Fern{\`{a}}ndez-Garcia, C.~V. Henri, D.~W. Hyndman,
  M.~S. Phanikumar, D.~Bolster, {Elimination of the Reaction Rate “Scale
  Effect”: Application of the Lagrangian Reactive Particle-Tracking Method to
  Simulate Mixing-Limited, Field-Scale Biodegradation at the Schoolcraft (MI,
  USA) Site}, Water Resources Research 53~(12) (2017) 10411--10432.
\newblock \href {https://doi.org/10.1002/2017WR021103}
  {\path{doi:10.1002/2017WR021103}}.

\bibitem{Rahbaralam2015}
M.~Rahbaralam, D.~Fern{\`{a}}ndez-Garcia, X.~Sanchez-Vila, {Do we really need a
  large number of particles to simulate bimolecular reactive transport with
  random walk methods? A kernel density estimation approach}, Journal of
  Computational Physics 303 (2015) 95--104.
\newblock \href {https://doi.org/10.1016/j.jcp.2015.09.030}
  {\path{doi:10.1016/j.jcp.2015.09.030}}.

\bibitem{Benson2016}
D.~A. Benson, D.~Bolster, {Arbitrarily complex chemical reactions on
  particles}, Water Resources Research 52~(11) (2016) 9190--9200.
\newblock \href {https://doi.org/10.1002/2016WR019368}
  {\path{doi:10.1002/2016WR019368}}.

\bibitem{Engdahl2017}
N.~B. Engdahl, D.~A. Benson, D.~Bolster, {Lagrangian simulation of mixing and
  reactions in complex geochemical systems}, Water Resources Research 53~(4)
  (2017) 3513--3522.
\newblock \href {https://doi.org/10.1002/2017WR020362}
  {\path{doi:10.1002/2017WR020362}}.

\bibitem{Herrera2009}
P.~A. Herrera, M.~Massab{\'{o}}, R.~D. Beckie,
  \href{http://linkinghub.elsevier.com/retrieve/pii/S0309170808002273}{{A
  meshless method to simulate solute transport in heterogeneous porous media}},
  Advances in Water Resources 32~(3) (2009) 413--429.
\newblock \href {https://doi.org/10.1016/j.advwatres.2008.12.005}
  {\path{doi:10.1016/j.advwatres.2008.12.005}}.
\newline\urlprefix\url{http://linkinghub.elsevier.com/retrieve/pii/S0309170808002273}

\bibitem{Sole-Mari2017}
G.~Sole-Mari, D.~Fern{\`{a}}ndez-Garcia, P.~Rodr{\'{i}}guez-Escales,
  X.~Sanchez-Vila, {A KDE-Based Random Walk Method for Modeling Reactive
  Transport With Complex Kinetics in Porous Media}, Water Resources
  Research\href {https://doi.org/10.1002/2017WR021064}
  {\path{doi:10.1002/2017WR021064}}.

\bibitem{Sole2018}
G.~Sole-Mari, D.~Fern\`andez-Garcia, Lagrangian modeling of reactive transport
  in heterogeneous porous media with an automatic locally adaptive particle
  support volume, Water Resources Research 54~(10) (2018) 8309--8331.
\newblock \href {https://doi.org/10.1029/2018WR023033}
  {\path{doi:10.1029/2018WR023033}}.

\bibitem{Szymczak2003}
P.~Szymczak, A.~J.~C. Ladd, Boundary conditions for stochastic solutions of the
  convection-diffusion equation, Phys. Rev. E 68 (2003) 036704.
\newblock \href {https://doi.org/10.1103/PhysRevE.68.036704}
  {\path{doi:10.1103/PhysRevE.68.036704}}.

\bibitem{Szymczak2004}
P.~Szymczak, A.~J.~C. Ladd, Stochastic boundary conditions to the
  convection-diffusion equation including chemical reactions at solid surfaces,
  Phys. Rev. E 69 (2004) 036704.
\newblock \href {https://doi.org/10.1103/PhysRevE.69.036704}
  {\path{doi:10.1103/PhysRevE.69.036704}}.

\bibitem{Koch2014}
J.~Koch, W.~Nowak, A method for implementing dirichlet and third-type boundary
  conditions in ptrw simulations, Water Resources Research 50~(2)  1374--1395.
\newblock \href
  {http://arxiv.org/abs/https://agupubs.onlinelibrary.wiley.com/doi/pdf/10.1002/2013WR013796}
  {\path{arXiv:https://agupubs.onlinelibrary.wiley.com/doi/pdf/10.1002/2013WR013796}},
  \href {https://doi.org/10.1002/2013WR013796}
  {\path{doi:10.1002/2013WR013796}}.

\bibitem{Boccardo2018}
G.~Boccardo, I.~M. Sokolov, A.~Paster, An improved scheme for a robin boundary
  condition in discrete-time random walk algorithms, Journal of Computational
  Physics 374 (2018) 1152 -- 1165.
\newblock \href {https://doi.org/https://doi.org/10.1016/j.jcp.2018.08.009}
  {\path{doi:https://doi.org/10.1016/j.jcp.2018.08.009}}.

\bibitem{Fernandez-Garcia2011}
D.~Fern\`andez-Garcia, X.~Sanchez-Vila, {Optimal reconstruction of
  concentrations, gradients and reaction rates from particle distributions},
  Journal of Contaminant Hydrology 120-121~(C) (2011) 99--114.
\newblock \href {https://doi.org/10.1016/j.jconhyd.2010.05.001}
  {\path{doi:10.1016/j.jconhyd.2010.05.001}}.

\bibitem{Pedretti2013}
D.~Pedretti, D.~Fern{\`{a}}ndez-Garcia, {An automatic locally-adaptive method
  to estimate heavily-tailed breakthrough curves from particle distributions},
  Advances in Water Resources 59 (2013) 52--65.
\newblock \href {https://doi.org/10.1016/j.advwatres.2013.05.006}
  {\path{doi:10.1016/j.advwatres.2013.05.006}}.

\bibitem{Silverman1986}
B.~W. Silverman, {Density Estimation for Statistics and Data Analysis},
  Vol.~37, 1986.
\newblock \href {https://doi.org/10.2307/2347507} {\path{doi:10.2307/2347507}}.

\bibitem{botev2010}
Z.~I. Botev, J.~F. Grotowski, D.~P. Kroese, {Kernel density estimation via
  diffusion}, Ann. Statist. 38~(5) (2010) 2916--2957.
\newblock \href {https://doi.org/10.1214/10-AOS799}
  {\path{doi:10.1214/10-AOS799}}.

\bibitem{Marron_bound}
J.~S. Marron, D.~Ruppert, Transformations to reduce boundary bias in kernel
  density estimation, Journal of the Royal Statistical Society. Series B
  (Methodological) 56~(4) (1994) 653--671.

\bibitem{Sole2019SPH}
G.~Sole-Mari, M.~J. Schmidt, S.~D. Pankavich, D.~A. Benson, Numerical
  equivalence between sph and probabilistic mass transfer methods for
  lagrangian simulation of dispersion, Advances in Water Resources 126 (2019)
  108 -- 115.
\newblock \href
  {https://doi.org/https://doi.org/10.1016/j.advwatres.2019.02.009}
  {\path{doi:https://doi.org/10.1016/j.advwatres.2019.02.009}}.

\bibitem{stochasticproblems}
S.~Chandrasekhar, Stochastic problems in physics and astronomy, Rev. Mod. Phys.
  15 (1943) 1--89.
\newblock \href {https://doi.org/10.1103/RevModPhys.15.1}
  {\path{doi:10.1103/RevModPhys.15.1}}.

\bibitem{vanGenuchten1982}
M.~T. van Genuchten, W.~J. Alves, Analytical solutions of the one-dimensional
  convective-dispersive solute transport equation, Technical Bulletins 157268,
  United States Department of Agriculture, Economic Research Service (1982).

\bibitem{Boccardo_Robin_18}
G.~Boccardo, I.~M. Sokolov, A.~Paster, An improved scheme for a robin boundary
  condition in discrete-time random walk algorithms, Journal of Computational
  Physics 374 (2018) 1152 -- 1165.
\newblock \href {https://doi.org/https://doi.org/10.1016/j.jcp.2018.08.009}
  {\path{doi:https://doi.org/10.1016/j.jcp.2018.08.009}}.

\bibitem{HarbaughArlen2005}
A.~W. Harbaugh, {MODFLOW-2005 , The U . S . Geological Survey Modular
  Ground-Water Model — the Ground-Water Flow Process}, U.S. Geological Survey
  Techniques and Methods (2005) 253\href {https://doi.org/U.S. Geological
  Survey Techniques and Methods 6-A16} {\path{doi:U.S. Geological Survey
  Techniques and Methods 6-A16}}.

\bibitem{Bear2010}
J.~Bear, A.~Cheng, {Modeling Groudwater flow and contaminant transport}, in:
  Theory and Applications of Transport in Porous Media, 2010, p. 850.
\newblock \href {http://arxiv.org/abs/9809069v1} {\path{arXiv:9809069v1}},
  \href {https://doi.org/10.1007/978-1-4020-6682-5}
  {\path{doi:10.1007/978-1-4020-6682-5}}.

\bibitem{LaBolle1996}
E.~M. LaBolle, G.~E. Fogg, A.~F.~B. Tompson, {Random-walk simulation of
  transport in heterogeneous porous media: Local mass-conservation problem and
  implementation methods}, Water Resources Research 32~(3) (1996) 583--593.
\newblock \href {https://doi.org/10.1029/95WR03528}
  {\path{doi:10.1029/95WR03528}}.

\bibitem{baks}
G.~Sole-Mari, {Bounded Adaptive Kernel Smoothing (BAKS) [Software]
  \phantom{,}}\href {https://doi.org/10.5281/zenodo.2762791}
  {\path{doi:10.5281/zenodo.2762791}}.

\end{thebibliography}

\end{document}